\pdfoutput=1
\documentclass[a4paper,10pt,amsmath,amssymb,nofootinbib,twocolumn,superscriptaddress,floats,longbibliography,aps,prd]{revtex4-2}
\usepackage[british]{babel}
\usepackage{txfonts,mathrsfs,bm}
\usepackage{natbib,textcase}
\usepackage{graphicx,color,soul}
\usepackage[pdftex,pdfusetitle]{hyperref}
\hypersetup{colorlinks=true,linkcolor=red,citecolor=blue,filecolor=black,urlcolor=black,
pdfauthor={Edgardo Franzin, Stefano Liberati, Jacopo Mazza, Ramit Dey and Sumanta Chakraborty}}
\usepackage[capitalize]{cleveref}
\crefname{subequation}{Eqs.}{Eqs.}
\labelcrefformat{subequation}{#2(#1)#3}
\usepackage[caption=false]{subfig}
\usepackage{physics}
\usepackage{array,booktabs} %for tables
\usepackage{multirow} %for tables
\makeatletter\g@addto@macro\bfseries{\boldmath}\makeatother%

\allowdisplaybreaks

\def\be#1\ee{\begin{align}#1\end{align}}

\newcommand{\ie}{i.e.}
\newcommand{\eg}{e.g.}
\renewcommand{\dd}{\text{d}}
\newcommand{\e}{\text{e}}

\newcommand{\0}{\nonumber}
\newcommand{\iu}{\text{i}}

\renewcommand{\geq}{\geqslant}

\renewcommand{\leq}{\leqslant}

\begin{document}

\title{\texorpdfstring{Scalar perturbations around rotating regular black holes and wormholes:\\quasi-normal modes, ergoregion instability and superradiance}{Scalar perturbations around rotating regular black holes and wormholes: quasi-normal modes, ergoregion instability and superradiance}}
\date{\today}

\newcommand{\SISSA}{\affiliation{SISSA, International School for Advanced Studies, via Bonomea 265, 34136 Trieste, Italy}}
\newcommand{\InfnTS}{\affiliation{INFN, Sezione di Trieste, via Valerio 2, 34127 Trieste, Italy}}
\newcommand{\IFPU}{\affiliation{IFPU, Institute for Fundamental Physics of the Universe, via Beirut 2, 34014 Trieste, Italy}}
\newcommand{\IACS}{\affiliation{School of Physical Sciences,
Indian Association for the Cultivation of Science, Kolkata-700032, India}}

\author{Edgardo Franzin}%\email{efranzin@sissa.it}
\author{Stefano Liberati}%\email{liberati@sissa.it}
\author{Jacopo Mazza}%\email{jmazza@sissa.it}
\SISSA\IFPU\InfnTS

\author{Ramit Dey}
\author{Sumanta Chakraborty}
\IACS

\begin{abstract}
We study scalar test-field perturbations on top of a Kerr--black-bounce background, \ie\ a family of rotating regular black holes and/or rotating traversable wormholes that can mimic Kerr black holes.
We compute the quasi-normal modes for a massless field in both the regular black holes and wormhole branches, confirming the stability of the former and identifying a set of growing modes that renders the latter unstable.
We further compute the superradiance amplification factors, for massless and massive fields, in the regular black hole branch, confirming that these objects superradiate, though to a lesser degree than the corresponding Kerr black holes. 
\end{abstract}

\maketitle%

\section{Introduction}

Over the past century, general relativity (GR) and black holes have been extensively tested with great success.
In particular, the recent detection of gravitational waves from the coalescence of binary compact objects~\cite{LIGOScientific:2016aoc,LIGOScientific:2018mvr,LIGOScientific:2020ibl,LIGOScientific:2021djp} and the observation of the shadow cast by the supermassive body at the center of M87~\cite{EventHorizonTelescope:2019dse} were huge leaps in the direction of getting a direct confirmation about the existence of black holes and thus testing the robustness of GR\@.
Nonetheless, the existence of singularities in GR and the inability to resolve them within the classical framework of the theory indicates the breakdown/incompleteness of the theory at high energy scales (typically in the Planckian regime) and in particular at black hole cores.
While it is widely believed that quantum effects may cure singularities, we currently lack a consistent UV-complete quantum theory of gravity to settle the issue.
However, we can still gain relevant insights by postulating different quantum gravity scenarios and studying the corresponding regularized effective spacetime geometries stemming from them. Indeed, current observational bounds/constraints on compact objects leave room for studying and speculating about some of these geometries as black-hole alternatives/mimickers~\cite{Damour:2007ap,Cardoso:2016rao}.
To this end, a large number of models --- inspired by quantum gravity scenarios and other arguments --- have been proposed over the years. Only few of them, however, represent viable outcomes of gravitational collapse~\cite{Carballo-Rubio:2019nel,Carballo-Rubio:2019fnb}.
Two examples in this category are regular black holes (trapped regions characterized by a regular core) and wormholes. Instances of both these types of compact objects are commonly seen in the literature (particularly within semi-classical or quantum theories of gravity)~\cite{Simpson:2018tsi}, and for this reason the research focus is now shifting more towards testing whether these alternative geometries give rise to characteristic signatures albeit closely mimicking the typical features of classical GR black holes.

After the release of the Event Horizon Telescope Collaboration's picture of the central massive object of M87~\cite{EventHorizonTelescope:2019dse}, considerable effort has been put into computing shadows of well-motivated candidate ultracompact objects belonging to theories alternative to GR~\cite{Banerjee:2019nnj}. In order to probe other plausible signatures of these black-hole mimickers, the computation of the quasi-normal modes (QNMs) has also been the subject of intense scrutiny. Any deviation from the classical black-hole picture is expected to have an imprint on the QNM spectrum. Since the post-merger ringdown of a classical black hole is described in terms of the QNMs, this opens up the possibility of testing any deviations from GR by analyzing the ringdown signals as obtained by LIGO and Virgo~\cite{Cardoso:2016oxy,Carson:2019kkh,Abedi:2020ujo,Bhagwat:2021kfa}, or in the future by the proposed third generation detectors~\cite{Maggiore:2019uih}.
QNMs are computed within the perturbative regime of the background theory and are sensitive to the boundary conditions imposed at asymptotic infinity as well as on the horizon~\cite{Oshita:2019sat,Abedi:2018npz,Dey:2020lhq}.
Hence any modification to the near-horizon geometry, such as in the case of the exotic compact objects~\cite{Bueno:2017hyj,Mark:2017dnq}, where the event horizon is removed due to quantum effects (or, exotic matter fields), as well as any deviations from the well-studied Kerr geometry can be directly linked to gravitational-wave observations via the QNM analysis~\cite{Konoplya:2016pmh}. 

Besides studying the QNM spectrum and the shadow measurements for these black-hole mimickers (or, few other candidate spacetimes), as well as looking for deviations from GR, it is also important to explore the stability of these objects under small perturbations.
In particular, if they are rotating (mimicking a Kerr black hole) and possess an ergosphere, one must carefully address the issues arising due to superradiant instability~\cite{Press:1972zz,Cardoso:2004nk,Hod:2017cga,Addazi:2019bjz,brito_superradiance_2020,Franzin:2021kvj}.
For example, a highly reflective as well as rapidly rotating black-hole mimicker would suffer from superradiant instability~\cite{Cardoso:2007az,Cardoso:2008kj,Pani:2010jz,Maggio:2017ivp,Maggio:2018ivz,Dey:2020pth}.
This is because, the amplified (superradiant) modes get further amplified due to reflection from the reflective surface in the near-horizon region of the black-hole mimicker and their repeated passage through the ergoregion, while for black holes all these amplified modes would have been absorbed by the event horizon, thus taming superradiant instability.
Therefore it is of utmost importance to study the phenomenon of superradiance for rotating black-hole mimickers, in order to understand their viability. 

It is worth stressing that among the possible black hole mimickers, regular black holes are further plagued by another kind of instability, linked to the required existence of an inner horizon. Indeed, these structures typically suffer from the so-called ``mass inflation'' instability~\cite{Carballo-Rubio:2018pmi, Carballo-Rubio:2021bpr} which renders them at most meta-stable solutions possibly leading to other regular geometries~\cite{Carballo-Rubio:2019nel,Carballo-Rubio:2019fnb} (if one postulates that quantum gravity will always avoid the formation of a singularity). 
This mechanism (which also applies in the presence of a cosmological constant~\cite{DiFilippo:2020ooa}) cuts short the ongoing debate about the survival of the strong cosmic censorship conjecture~\cite{Cardoso:2017soq,Dias:2018ynt,Mishra:2020jlw,Rahman:2020guv,Rahman:2018oso}, but challenges the regular black-hole scenarios as a viable resolution of singularities.

Also in order to side step this issue, we shall focus here on the Kerr--black-bounce scenario~\cite{Mazza:2021rgq}, a family of regular black holes and/or traversable wormholes capable of mimicking Kerr black holes to an impressive extent. In particular, these solutions, in the black-hole case, are regularized by a wormhole throat which can be large enough to avoid the presence of an inner horizon and hence its associated mass inflation instability. 
Notably, these geometries are also simple in that they are described by a single additional parameter, other than the mass and the spin, that regularizes the singularity and basically describe the size of the wormhole throat. Indeed, this class of compact objects interpolates smoothly between regular black holes and traversable wormholes, depending on the choice of the spin and this regularizing parameter.
For all these reasons, Kerr--black-bounce solutions recently received considerable attention~\cite{bambhaniya_thin_2021,guerrero_shadows_2021,Islam:2021ful,jafarzade_observational_2021,lima_junior_mistaken_2021,shaikh_constraining_2021} and we here contribute to the study of their phenomenology by investigating the dynamics of a scalar test field propagating on top of the Kerr--black-bounce background.
Specifically, we consider a massless scalar field and compute the QNMs; when the background is a traversable wormhole, we further search for unstable modes and derive the ensuing instability timescale. Finally, we endow the field with a mass and study the superradiance by computing the amplification factors.

The paper is organized as follows: We start by reviewing the basic aspects of the Kerr--black-bounce scenario in \cref{sec:background}. Perturbations of this background due to a scalar test field are introduced in \cref{sec:scalar}, while specification to the QNM case is deferred to \cref{sec:QNM}, in which the QNM frequencies associated with the Kerr--black-bounce scenario are determined. We study the spectrum of superradiant amplification in \cref{sec:Superradiance}, then we conclude in the subsequent section. Hereafter we have set the fundamental constants $G=c=1$.

\section{Background metric\label{sec:background}}

In this section we briefly review the background spacetime, which is described by the Kerr--black-bounce metric~\cite{Mazza:2021rgq}:
\be\label{background_orig}
\dd s^2 = &-\left(1-\frac{2M \sqrt{r'^2 + \ell^2}}{\Sigma} \right) \dd t^2 + \frac{\Sigma}{\Delta}\,\dd r'^2 + \Sigma\,\dd \theta^2\0\\
&- \frac{4M a \sin^2 \theta \sqrt{r'^2 + \ell^2}}{\Sigma}\,\dd t\,\dd \varphi + \frac{A \sin^2\theta}{\Sigma}\,\dd \varphi^2 
\ee
with $M$ and $a$ being the mass and the spin of the spacetime, and $\ell$ a real positive regularizing parameter, with
\be
\Sigma &= r'^2 + \ell^2 + a^2 \cos^2\theta, \quad
\Delta = r'^2 + \ell^2 - 2M \sqrt{r'^2 + \ell^2} + a^2, \0\\
A &= \left(r'^2 + \ell^2 + a^2\right)^2 - \Delta a^2 \sin^2\theta\,.
\ee
Note that, in the limit $\ell \rightarrow 0$, the Kerr--black-bounce metric reduces to the Kerr metric.
The above line element effectively adds rotation to the Simpson--Visser black-bounce metric~\cite{Simpson:2018tsi, simpson_vaidya_2019, lobo_novel_2021} through the Newman--Janis algorithm and has been recently extended to charged spacetimes~\cite{Franzin:2021pyi}.

The coordinates $(t,r',\theta,\varphi)$ are convenient to classify the spacetime. According to the value of the parameter $\ell$, the line element in \cref{background_orig} describes a regular black hole or a traversable wormhole.
Notice that $r'$ may take negative values as well, in the sense that the metric is symmetric under the exchange $r'\to -r'$, meaning that the spacetime describes two identical patches glued at $r'=0$; we will refer to the two patches as ``our universe'', for $r'>0$, and the ``other universe'', for $r'<0$.
When $\ell\neq0$, the spacetime is free of singularities and $r'=0$ represents a regular finite traversable surface, \ie\ $r'=0$ represents a wormhole throat, whose nature (timelike, null or spacelike) depends on the specific values of $a$ and $\ell$.
The line element presented in \cref{background_orig} may have coordinate singularities for values of $r'$ such that $\Delta=0$, which turn out to be the event horizons,
\be\label{horizons_orig}
r'_\pm = \sqrt{r_\pm^2 - \ell^2}\,,\quad
r_\pm \equiv M\pm\sqrt{M^2-a^2}\,.
\ee
Thus, depending on the choice of the parameters, we may have two (if $a<M$ and $\ell<r_-$), one (if $a<M$ and $r_-<\ell<r_+$) or no coordinate singularities (if $a<M$ and $\ell>r_+$, or if $a>M$).
Note, in particular, that for $r_- < \ell < r_+$, the regular black hole has no inner horizons; as mentioned in the Introduction, this is quite an attractive feature --- one not shared by most regular black holes described in the literature --- as it entails this spacetime might avoid mass inflation. 
The complete classification, including the limiting cases, together with the corresponding Penrose diagrams, can be found in Ref.~\cite{Mazza:2021rgq}.
Here, we will only distinguish between configurations with $\ell< r_+$, which we will call regular black holes, and $\ell> r_+$, to which we will refer to as (traversable) wormholes.
The intermediate case $\ell=r_+$, corresponding to a wormhole whose throat is null and coincides with the (extremal) event horizon, will often require specific considerations.

For the scope of this paper, it is better to perform the coordinate transformation to a new radial coordinate $r=\sqrt{r'^2+\ell^2}$ and work with the metric
\be\label{background}
\dd s^2 = &-\left(1-\frac{2Mr}{\Sigma} \right) \dd t^2 + \frac{\Sigma}{\delta\Delta}\,\dd r^2 + \Sigma\,\dd \theta^2\0\\
&- \frac{4M a r \sin^2\theta}{\Sigma}\,\dd t\,\dd\varphi + \frac{A \sin^2\theta}{\Sigma}\,\dd \varphi^2 
\ee
with now
\be
\Sigma &= r^2 + a^2 \cos^2\theta, \quad
\Delta = r^2 - 2Mr + a^2, \quad
\delta = 1-\frac{\ell^2}{r^2},\0\\
A &= \left(r^2 + a^2\right)^2 - \Delta a^2 \sin^2\theta\,.
\ee
With these coordinates $(t,r,\theta,\phi)$ we recognize the metric in \cref{background} as a particular case of the Johannsen family~\cite{Johannsen:2015pca,Johannsen:2015mdd}.
Note that now $r\geq \ell$, with $r=\ell$ representing the wormhole throat; the horizons, if any, are located at $r=r_\pm$.

We stress here that the line element in \cref{background_orig}, or alternatively in \cref{background}, is motivated by quantum gravity arguments and is not a general relativistic solution.
Yet, it is reasonable to think that, once the configuration settles down and becomes stationary, quantum gravity effects will be accountable in terms of an effective stress-energy tensor.
This effective stress-energy tensor is proportional to the Einstein tensor and describes the matter content of the solution.
Details can be found in Ref.~\cite{Mazza:2021rgq}, but the important fact is that the matter content of the spacetime is localized close to the origin $r'=0$ (\ie\ $r=\ell$) and energy density and pressures fall off as $1/r^4$.
This means that the spacetime is effectively vacuum even close to the throat and hence the line element in \cref{background_orig} describes a good black-hole mimicker.

\section{Scalar perturbations\label{sec:scalar}}

In this section we will study the perturbation on the background Kerr--black-bounce geometry due to a test scalar field. For this purpose, we start with the Klein--Gordon equation $\Box\phi=\mu^2\phi$ for a massive scalar field $\phi$ with mass $m_\phi=\hbar\mu$. Further, assuming the decomposition $\phi=\e^{\iu m \varphi}\e^{-\iu\omega t}S(\theta)R(r)$, with $m$ and $\omega$ being the azimuthal number and the frequency of the perturbation, the Klein--Gordon equation separates into an angular equation 
\be\label{angulareq}
\frac{1}{\sin\theta}\frac{\dd}{\dd\theta}&\left(\sin\theta\frac{\dd S}{\dd\theta}\right) \0\\
+ &\left(a^2 \left(\omega^2-\mu^2\right)\cos^2\theta + A_{lm} - \frac{m^2}{\sin^2\theta}\right)S = 0\,,
\ee
\ie\ the spheroidal harmonics equation, whose eigenvalues $A_{lm}$ are also characterized by the harmonic number $l$, and a radial equation
\be\label{radialeq}
\sqrt{\delta}\,\frac{\dd}{\dd r} \left(\sqrt{\delta}\Delta \frac{\dd R}{\dd r}\right) + \left(\frac{K^2}{\Delta} - \mu^2 r^2 - \lambda\right)R = 0\,,
\ee
where $K = \left(r^2+a^2\right)\omega - am$ and $\lambda=A_{lm} - 2am\omega + a^2 \omega^2$.

In the non-rotating limit, \cref{angulareq} reduces to the spherical harmonics equation with eigenvalues $A_{lm}=l(l+1)$. More generally, \cref{angulareq} must be solved perturbatively in $a\omega$ or numerically~\cite{Berti:2005gp}.
In our computations, we have evaluated the angular eigenvalue both numerically with the Leaver method~\cite{Leaver:1985ax} and approximately with a high-order expansion in $a\omega$.

For the radial equation, on the other hand, two limits are worth considering: one corresponding to spatial infinity, \ie\ $r\rightarrow \infty$, and one to the near-horizon or near-throat region, depending on the background geometry. %See \cref{radial_asymp} for further details.

At spatial infinity, the radial function has the following asymptotic behavior
\be\label{Rasympmass}
R(r) \sim \frac{1}{r}\,\e^{q r} r^{M \left(\mu^2-2 \omega^2\right)/q}\,,\quad
q=\pm\sqrt{\mu^2-\omega^2}\,.
\ee
The sign of the real part of $q$ determines the behavior of the wavefunction at $r\to\infty$.
If $\Re(q)>0$ the solution diverges, while for $\Re(q)<0$ the solution tends to zero. The general solution will be a linear combination of both cases.

In the massless case, \cref{Rasympmass} reduces to a simpler form,
\be\label{Rasympmassless}
R(r) \sim \frac{1}{r}\,\e^{\pm\iu\omega r} r^{\pm 2\iu M\omega}\,
\ee
where the plus (minus) sign corresponds to outgoing (ingoing) waves.
It is to be noted that the asymptotic solution at spatial infinity is independent of the parameter $\ell$, but for determining the near-horizon or near-throat asymptotic solution, $\ell$ would play an important role, which we explore now. 

When the regularizing parameter $\ell$ satisfies $\ell<r_+$, the metric in \cref{background} describes a regular black hole and the two independent solutions close to the event horizon behave as
\be\label{leading_bh}
R(r) \sim (r-r_+)^{\pm \iu\sigma}\,,
\quad \sigma = \frac{a m-2 M \omega r_+}{\gamma\left(r_+-r_-\right)}\,,
\quad \gamma = \sqrt{1-\frac{\ell^2}{r_+^2}} \,.
\ee

For traversable wormholes with regularizing parameter $\ell>r_+$, close to the throat the two linearly independent solutions are asymptotic to
\be\label{leading_woh}
R(r) &\sim \exp \left(\pm\frac{\iu \tilde{\omega} \left(a^2+\ell^2\right) \sqrt{2\ell (r-\ell)}}{\Delta (\ell)}\right),\0\\
\tilde{\omega}^2 &= \left(\omega -\frac{a m}{a^2+\ell^2}\right)^2-\frac{\Delta (\ell) \left(\ell^2 \mu ^2+\lambda \right)}{\left(a^2+\ell^2\right)^2}-\frac{\Delta (\ell)^2}{\left(a^2+\ell^2\right)^3}\,,
\ee
where $\Delta(\ell)$ means $\Delta$ evaluated at $r=\ell$.

In the particular case in which the throat of the wormhole becomes a null surface and coincides with the black-hole horizon, \ie\ for $\ell=r_+$, the corresponding solutions are of the form
\be\label{leading_nwoh}
R(r) \sim \exp \left(\pm \iu \frac{am-2M\omega r_+}{r_+ - r_-}\sqrt{\frac{2\ell}{r-\ell}}\right).
\ee

In \cref{leading_bh,leading_woh,leading_nwoh} the plus (minus) sign corresponds to outgoing (ingoing) waves.

\subsection{Boundary conditions\label{subsec:bc}}

For determining the QNMs or the superradiant amplification factors, one needs to supplement \cref{radialeq} with appropriate boundary conditions. Such boundary conditions define the physical problem at hand and depend on whether the spacetime contains a black hole or not. 

\begin{figure*}
\centering
\subfloat[The regular black hole. The maximally extended spacetime continues above and below the portion shown by repetition of this fundamental block.\label{fig:RBH}]{\includegraphics[width=0.3\textwidth]{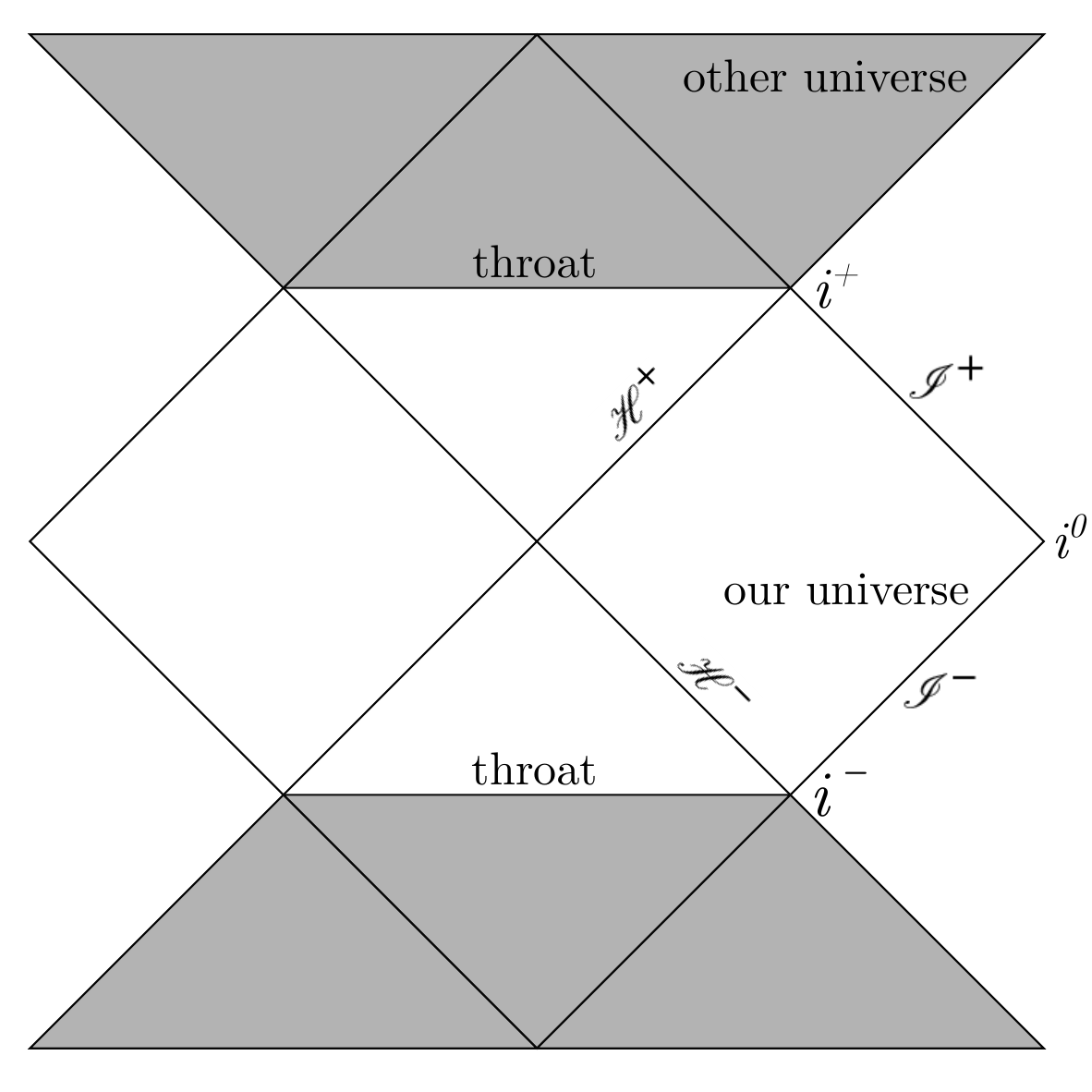}}
\hfill
\subfloat[The null-throat wormhole. The analytically extended spacetime continues above and below by repetition of this fundamental block.\label{fig:nWoH}]{\includegraphics[width=0.3\textwidth]{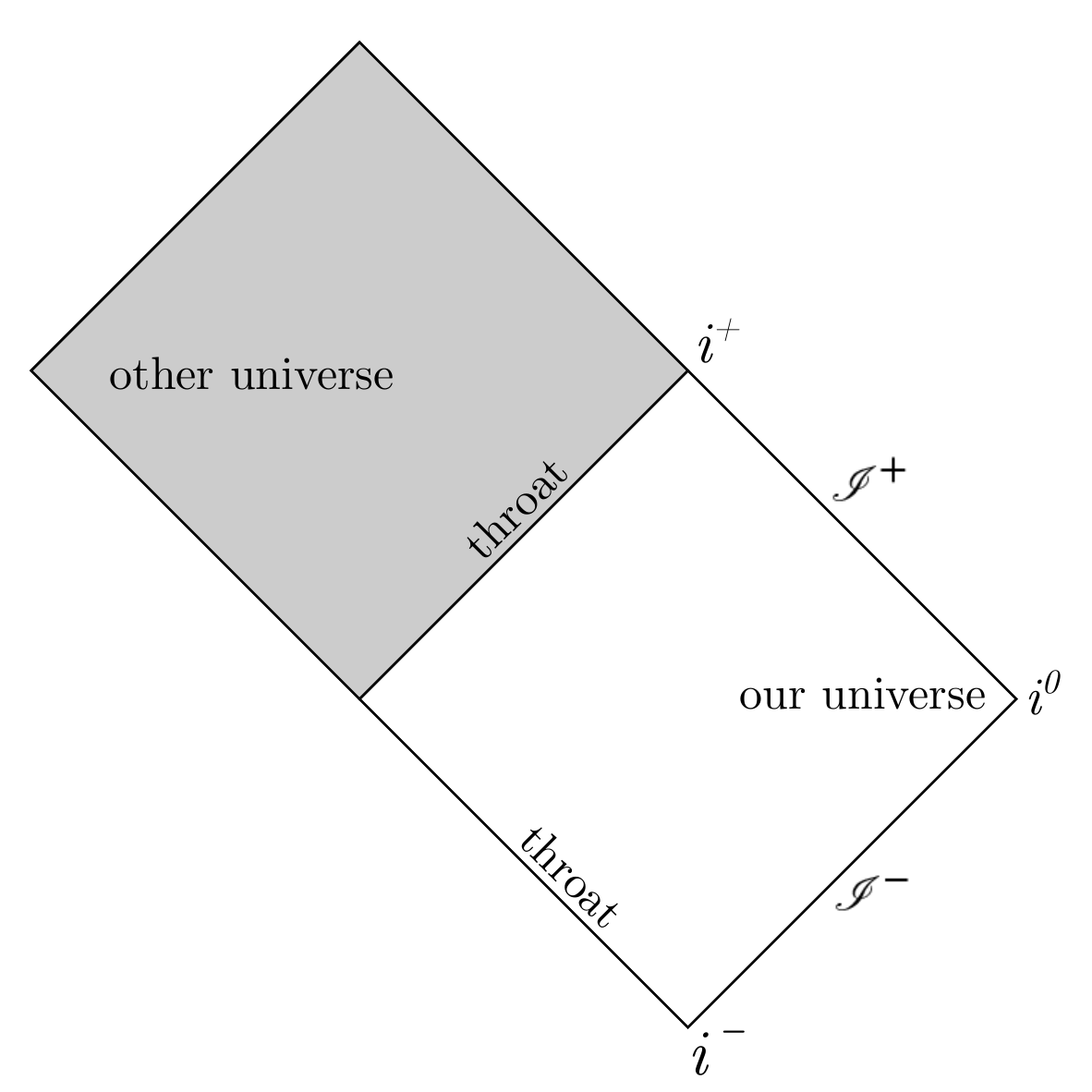}}
\hfill
\subfloat[The traversable wormhole.\label{fig:WoH}]{\includegraphics[width=0.3\textwidth]{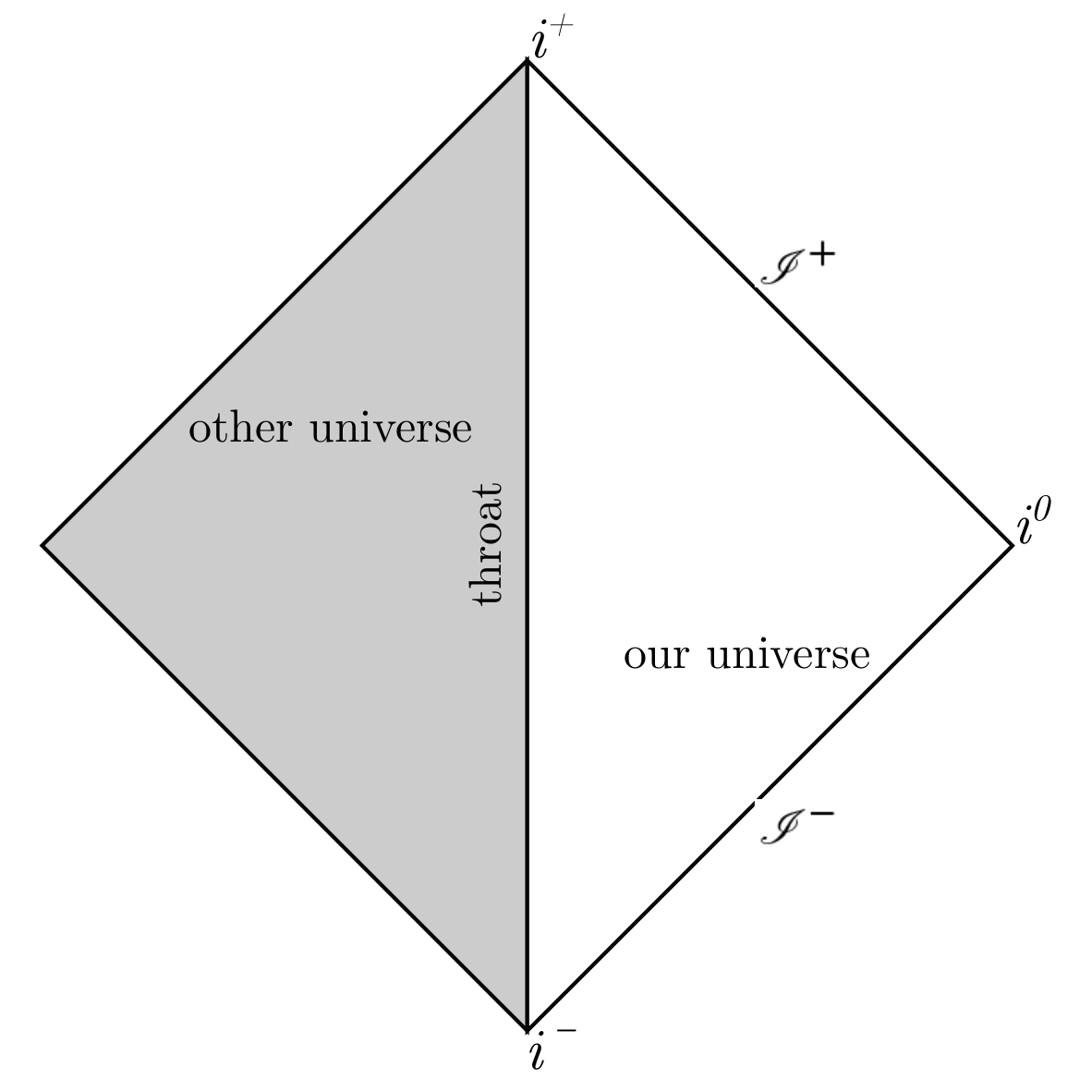}}
\caption{Penrose diagrams of regular black hole, null-throat wormhole and traversable wormhole. The white area represents ``our universe'' while the gray area is the ``other universe''.\label{fig:penrose}}
\end{figure*}

QNMs encode the scalar's late-time response to an initial perturbation that is localized in space. For this reason, we demand purely outgoing waves at spatial infinity. In the regular black hole case, we further demand that no radiation comes out of the horizon. The null throat case is analogous to the regular black hole, in this respect: As can be deduced by inspecting the conformal diagrams of \cref{fig:penrose}, in this case the wormhole throat coincides with the horizon and is therefore a causal boundary. (The only causal curves that reach $r=+\infty$ after having crossed $r=\ell$ originated from the ``other universe'' in the past analytical extension of the spacetime.) Hence, we impose purely ingoing boundary conditions at the null throat. When $\ell>r_+$, instead, the throat is traversable in both directions and the ``two universes'' are causally connected.
Since the geometry on the two sides of the wormhole is symmetric, we assume that the scalar field will inherit the symmetry of the background. This assumption translates into perfect reflection at the throat, which we implement by demanding $R(\ell) = 0$. Alternatively, one can require the derivative of the radial function to vanish at the throat; such Neumann boundary conditions are associated to another family of QNMs, whose computation is beyond the scope of this paper. 

Superradiance is, in essence, a scattering experiment whereby an ingoing wave is sent in from infinity, it scatters off the compact object and is then measured again at infinity.
As both ingoing and outgoing radiation must be present at spatial infinity, we allow for both solutions of \cref{leading_woh}.
For regular black holes and null wormholes, the conditions at the inner boundary (\ie\ at the horizon) are the same we impose for the QNMs computation.
In the traversable wormhole case, however, the assumption of perfect reflection at the throat is no longer justified.
Indeed, that assumption would entail that the same scattering experiment is performed simultaneously in the ``two universes''.
We rather resolve to study superradiance from the perspective of ``our universe'' alone, thus assuming no ingoing radiation at infinity in the ``other universe''; of course, it is possible to do otherwise, but that investigation lies beyond the scope of this work.
Under this circumstance, a simple argument --- which we report in \cref{sec:Superradiance} --- ensures that no supperradiant amplification can occur, regardless of the exact boundary conditions imposed at the throat. Our choice of boundary conditions is summarized in \cref{tab:boundcond}.

\begin{table}[ht]
\centering
\begin{tabular}[t]{ll}
\toprule
\multicolumn{2}{c}{\textbf{Inner boundary}} \\
\midrule
regular black hole ($\ell<r_+$) & pure absorption, cf.\ \cref{leading_bh} \\
\midrule
null-throat wormhole ($\ell=r_+$) & pure absorption, cf.\ \cref{leading_nwoh} \\
\midrule
traversable wormhole ($\ell>r_+$) & (QNMs) pure reflection, $R(\ell)=0$ \\
\bottomrule
\toprule
\multicolumn{2}{c}{\textbf{Infinity}, cf.\ \cref{Rasympmass,Rasympmassless}}\\
\midrule
QNMs & purely outgoing \\
\midrule
superradiance & ingoing and outgoing \\
\bottomrule
\end{tabular}
\caption{Behavior of the radial function close to the inner boundary, \ie\ the horizon for regular black holes and the throat for wormholes, and asymptotically, according to the physical problem under investigation.\label{tab:boundcond}}
\end{table}

Clearly, different/other choices of boundary conditions are possible. For instance, the symmetry between the ``two universes'' could easily be broken, \eg\ by the presence of some matter on one side of the wormhole but not the other; if this were the case, perfect reflection at the throat could not be justified. Alternatively, one could imagine a situation in which the background is still symmetric but the perturbation is not, as in a scattering problem whereby a wavepacket is prepared in ``our universe'' and sent towards the object; in such a case, boundary conditions at the throat might not be needed at all. Finally, one might envisage a scenario in which the exotic matter that keeps the wormhole open is not transparent to the perturbation; this would make the dynamics non-conservative even at the test-field level. 
All these possibilities, though interesting, lie beyond the scope of this work.

\section{Quasi-normal modes and (in)stability\label{sec:QNM}}

QNMs can be obtained by various analytical methods but the complicated form of the potential makes it difficult to solve the perturbation equation without added assumptions or imposing restrictions on the parameter space. In this section, we focus on obtaining the QNMs numerically by the direct integration and shooting techniques. This approach is valid both for the black-hole and the wormhole branches.
For the regular black holes, the QNMs can also be approximated using the more analytic WKB approach~\cite{Schutz:1985km,Iyer:1986np}, and its generalization to rotating backgrounds~\cite{Seidel:1989bp,Kokkotas:1991vz}.
Below we first detail the two methods, then present our results.

\subsection{Methods}

\subsubsection{Direct integration} 

The direct integration technique~\cite{Chandrasekhar:1975zza} works as follows.
First, consider the non-rotating case.
For the black hole, we integrate \cref{radialeq} supplied with the correct boundary conditions both from infinity and from the horizon to an intermediate point (typically the maximum of the scalar potential) and then we shoot for the value of $\omega$ such that the radial function and its derivative are continuous at the intermediate point.
The same procedure is followed for the null-throat wormhole, though this case is technically more subtle --- we provide more detail in \cref{appendix}.
For the wormhole, we only integrate from infinity and shoot for the value of $\omega$ such that the solution is zero at the throat. 

In practice, infinity is taken to be at some large value of $r$ --- \eg\ $75M$. Similarly, the integration must start or stop a small distance away from the horizon or the throat, since the coefficients of the differential equation diverge there. These parameters, along with the location of the intermediate point, are varied by small amounts in order to assess the stability of our numerical results, which are stable within a numerical accuracy of, typically, order $10^{-3}$ or less.
Moreover, shooting requires an initial guess for the value of the QNM frequency $\omega$.
In the black hole case, we looked for solutions in the vicinity of the tabulated value of the corresponding fundamental QNM of Kerr.
The wormhole case requires a more thorough mapping of the solutions to the eigenvalue problem.

For the rotating case, starting with a small value of $a/M$, we start by considering the non-rotating QNM frequencies as initial guess values and then we solve for the angular eigenvalue.
Next, we integrate the radial equation as in the non-rotating case and we shoot for the frequency $\omega$.
We repeat this procedure as long as the frequency $\omega$ converges to a constant value; in practice, this is often achieved within five iterations.
The QNM frequencies for configurations with higher values of $a/M$ are determined by using a previous frequency as initial guess and following their behavior as a function of the spin parameter.

\subsubsection{WKB} 

Alternatively, for regular black holes and null-throat wormholes, the QNMs can be determined with the WKB approach as well.

Let us begin with the non-rotating case again. In a nutshell, the WKB approximation connects two solutions in a matching region, and gives the best results when the matching region is around the maximum of the scalar potential, which in this case does not depend on the frequency of the perturbation.
Hence, the potential can be Taylor-expanded around the maximum of the potential and, at leading order, the QNM frequencies are given by
\be\label{basicWKB}
\omega^2 = V_0-\iu\sqrt{-2V''_0}\left(n+\frac{1}{2}\right),\quad n=0,1,\dots\,,
\ee
where a prime represents a derivative with respect to the tortoise coordinate, and the subscript ``0'' means evaluated at the maximum of the potential.
The integer $n$ is the overtone number and the QNM with $n=0$ is called the fundamental mode.
Higher order corrections to this equation have been computed, as well as approaches to increase its accuracy~\cite{Iyer:1986np,Iyer:1986nq,Konoplya:2003ii,Matyjasek:2017psv,Konoplya:2019hlu}.
In our computations, good agreement with the numerical results are achieved considering a fourth-order approximation. This is also motivated by the fact that for scalar perturbations in a Kerr background, especially for the lowest $l$ values, agreement of order 3\% with numerical results requires at least a fourth-order WKB approximation~\cite{Konoplya:2019hlu}.

The rotating case is more involved, as the scalar potential and the angular eigenvalues \emph{do} depend on the frequency. The strategy in this case is to work perturbatively in powers of $a\omega$.
For $a\omega$ sufficiently small, we expect to obtain good accuracy with this truncated series. In our computations, we have considered orders up to the sixth --- the highest for which analytical results are available. This choice allows us to explore intermediate values of the spin parameter.
The procedure to determine the QNM frequency is then, in essence, equivalent to the WKB method in the non-rotating case, and we need to numerically solve an equation of the form
\be\label{rotatingWKB}
\omega^2 = f\left(a,\omega,\ell,n,l,m\right),
\ee
in order to determine $\omega$, given $a$, $\ell$, $n$, $l$ and $m$.
Generically, \cref{rotatingWKB} will contain a number of spurious roots which we discard by starting with the well-defined solution for $a=0$ and following the roots for increasing $a/M$. 

\begin{figure*}
\includegraphics[width=0.19\textwidth]{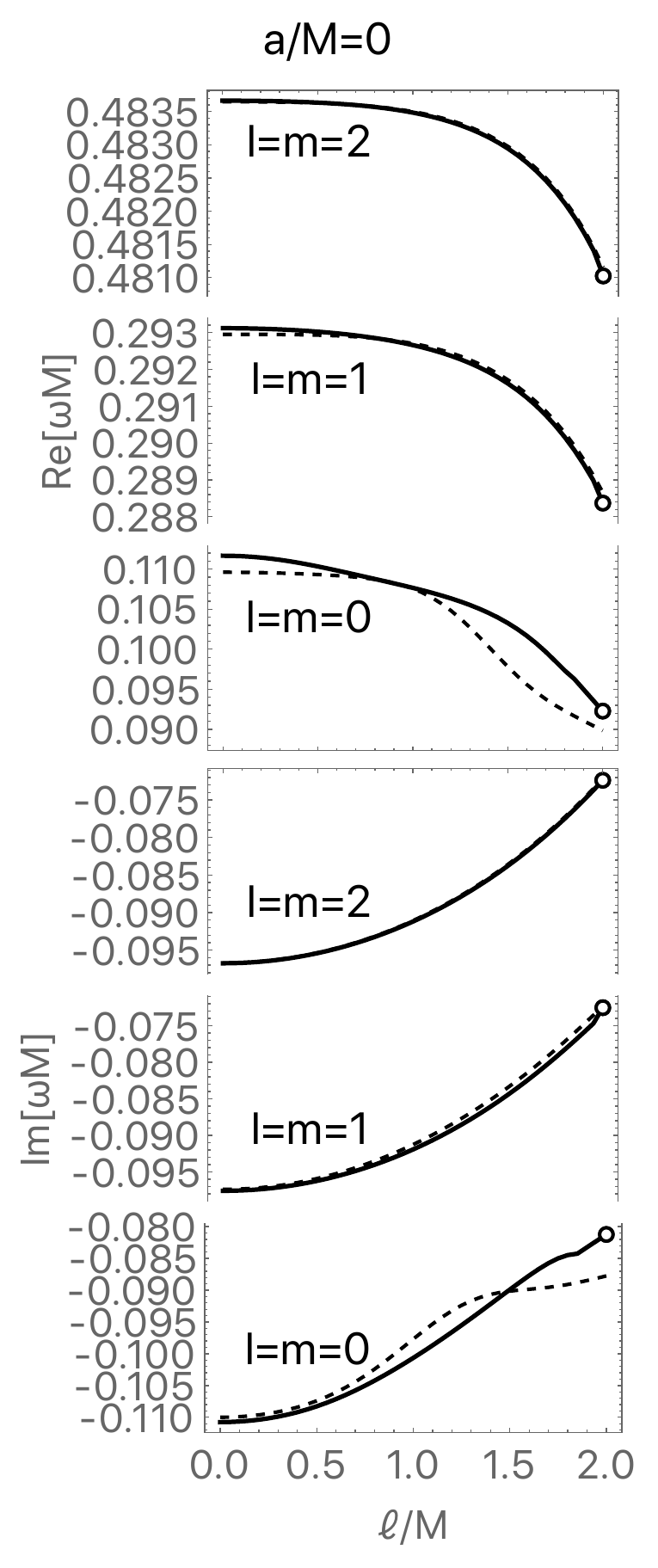}
\includegraphics[width=0.19\textwidth]{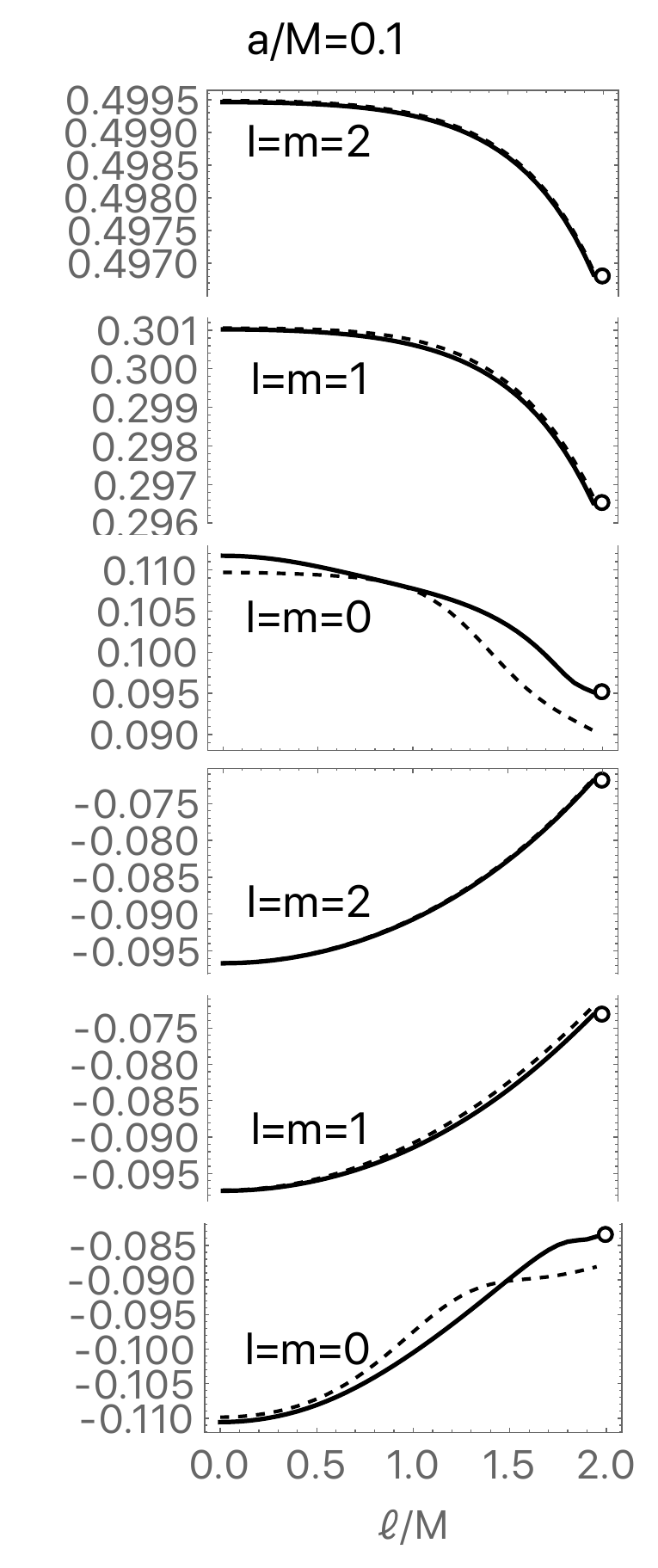}
\includegraphics[width=0.19\textwidth]{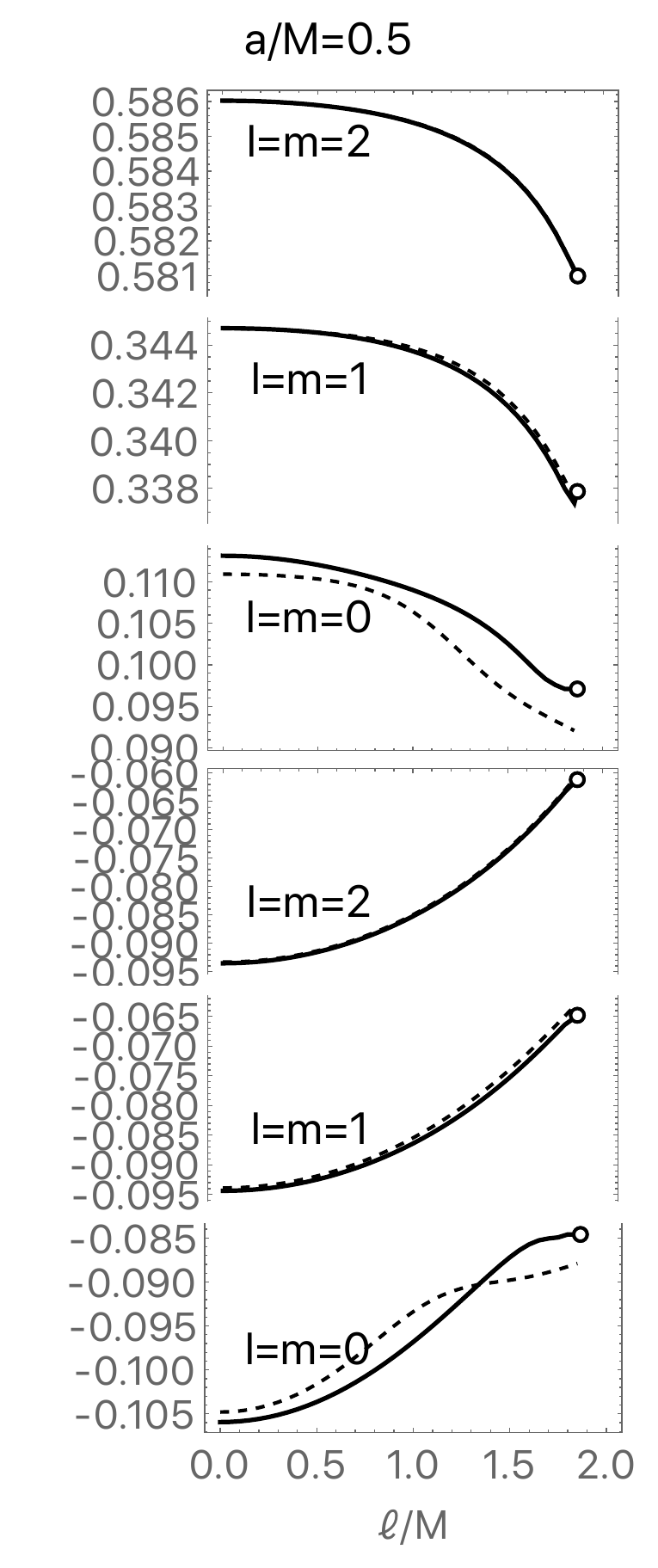}
\includegraphics[width=0.19\textwidth]{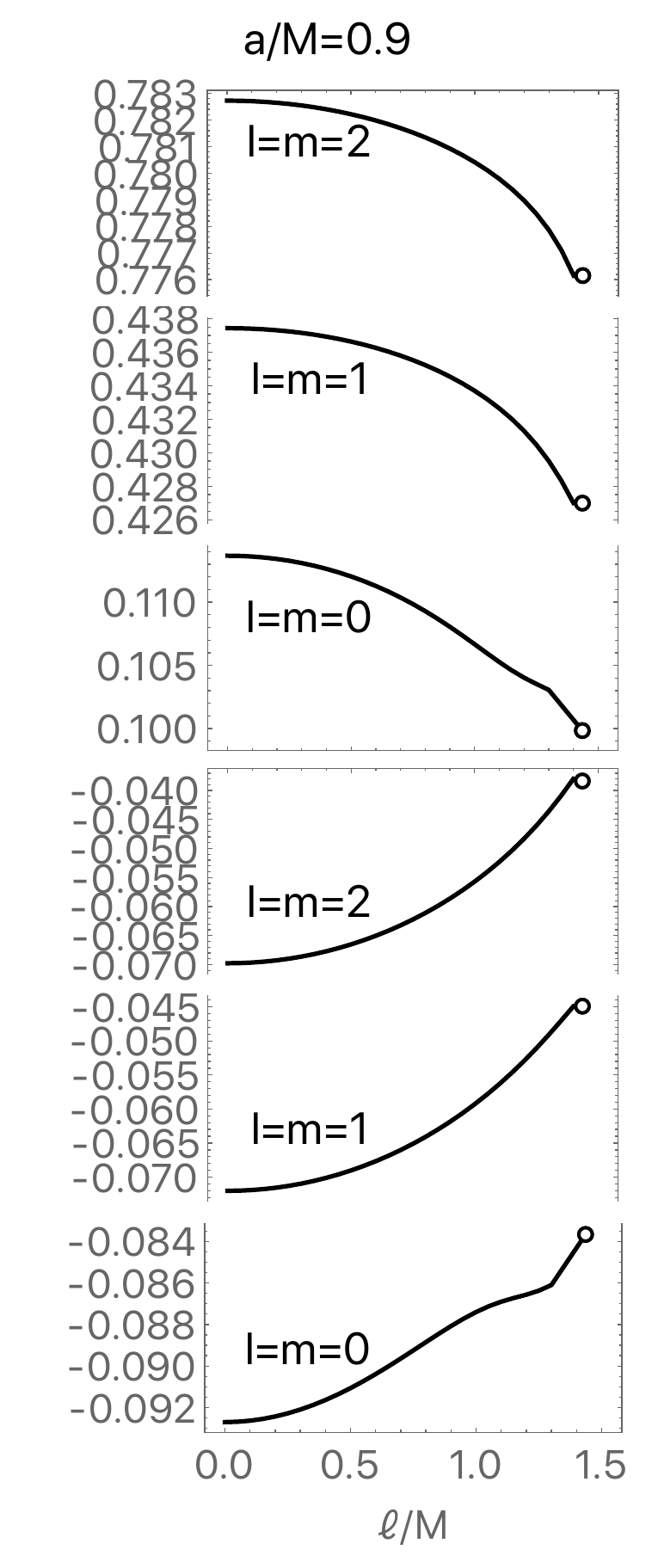}
\includegraphics[width=0.19\textwidth]{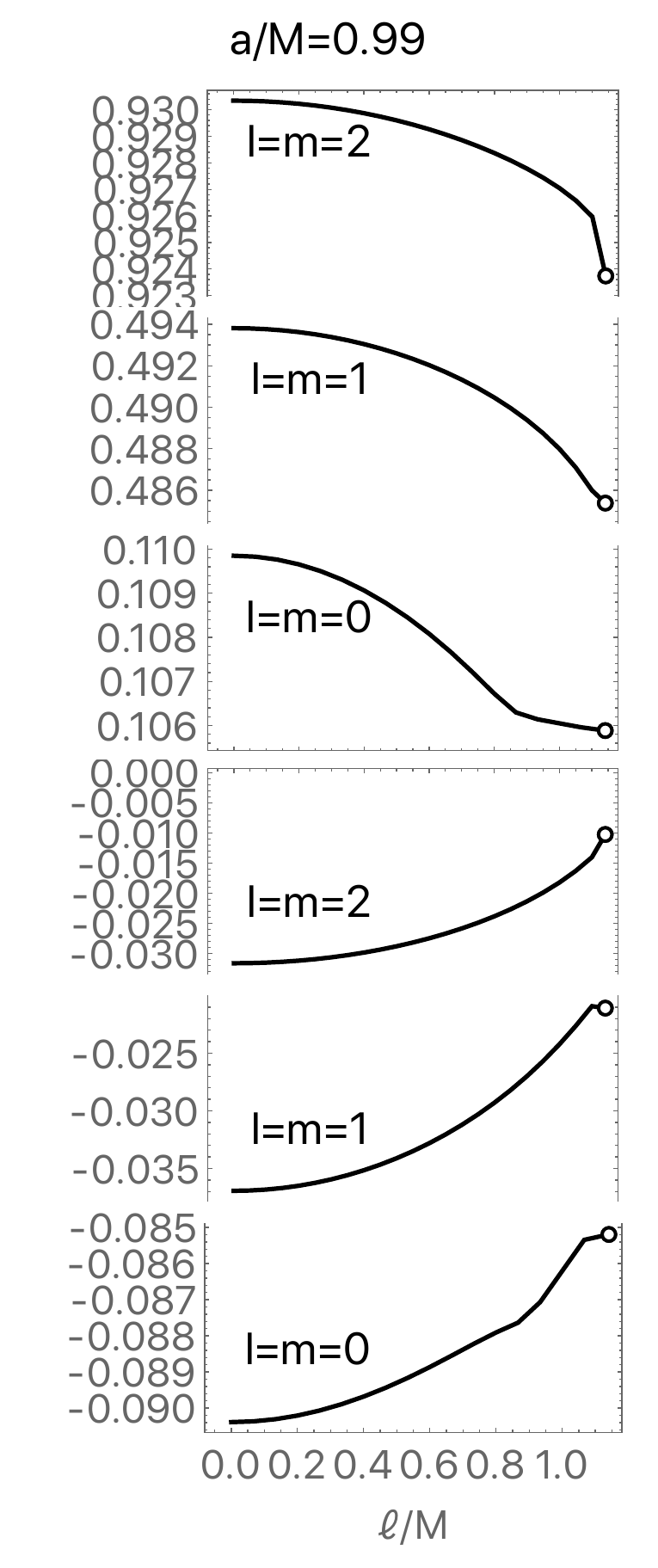}
\caption{QNMs for regular black holes and null-throat wormholes (empty circle for the $\ell=r_+$ case): real (top panels) and imaginary parts (bottom panels) of the QNM frequencies have been plotted as functions of the dimensionless regularizing parameter $(\ell/M)$, for the first few $l=m$ modes, for selected values of the spin parameter. The solid lines arise out of the direct integration scheme; while the dashed lines correspond to the WKB approximation, valid up to $a/M\lesssim0.5$.\label{fig:QNMsBH}}
\end{figure*}

\subsection{Results}

\subsubsection{Regular black holes}

Some of our results are presented in \cref{fig:QNMsBH}, where solid lines are obtained with the direct integration method, while dashed lines come from the WKB method. We verified that the results are not affected significantly by changes in the parameters entering our numerical routines (\ie\ the locations of the numerical infinity, of the numerical horizon and of the intermediate point). Clearly, the two methods are in good agreement for $a \lesssim 0.5 M$, although less so for the $l=m=0$ mode. This is not surprising, as the WKB approximation is expected to hold best for values of $l$ larger than the spin of the perturbation ($l>0$ in this case). In the non-rotating limit, our results are also in agreement with those in Ref.~\cite{Churilova:2019cyt}, obtained both with the WKB and time-domain methods. 
The fundamental QNM frequencies, as presented in \cref{fig:QNMsBH}, show a clear dependence on the regularizing parameter $\ell$, though the relative variations in their magnitude are rather mild. Each fundamental mode is accompanied by a whole tower of overtones which can in principle be computed with the same methods.

\subsubsection{Null-throat wormholes}

As mentioned before and as explained in detail in \cref{appendix}, the null-throat case is technically more involved than the regular black hole one.
The structure of the space of solutions is also more complex, as multiple modes (all stable) lie close to one another. As a result, we need higher accuracy in our numerical routines and very precise initial guesses for the shooting.
Otherwise, it is possible that small variations in the spacetime and integration parameters cause the numerical routine to ``jump'' between nearby modes, \eg\ from the fundamental mode to an overtone.

Despite the hurdles, a solid qualitative picture does emerge: wormholes with a null throat are stable, in the sense that their QNMs have negative imaginary part; and in all of the cases we have studied there exists a mode that can be reached along the curves of \cref{fig:QNMsBH}, the empty circle, in the limit $\ell \to r_+$\footnote{In same cases, the empty circle seems not to lie in the black-hole curve: this is due only to numerical precision.}.
In other words, wormholes with a null throat seem to be phenomenologically akin to regular black holes, and a limiting case thereof, in all respects hereby considered.

\subsubsection{Wormholes}

Given the different boundary conditions, there is no reason to expect that the curves of \cref{fig:QNMsBH} will cross over smoothly to the wormhole branch. Lacking guidance from known results, the frequency space had to be spanned more broadly in order to confidently identify the QNMs. More specifically, we considered a rectangular grid of points in the $\Im(\omega)$-$[\Re(\omega)>0]$ space, wide enough to enclose our rough expectations for the QNM frequency, and computed the quantity $\arg[ R(\ell)]$. A plot of this quantity permits to visually locate the zeroes of $R(\ell)$ in the frequency space, since the argument yields a recognizable pattern around them\footnote{This is analogous to what the Mathematica's \texttt{ComplexPlot} function does.}. In this way, we were able to pick accurate guesses for our shooting routine.

\begin{figure*}
\includegraphics[width=0.19\textwidth]{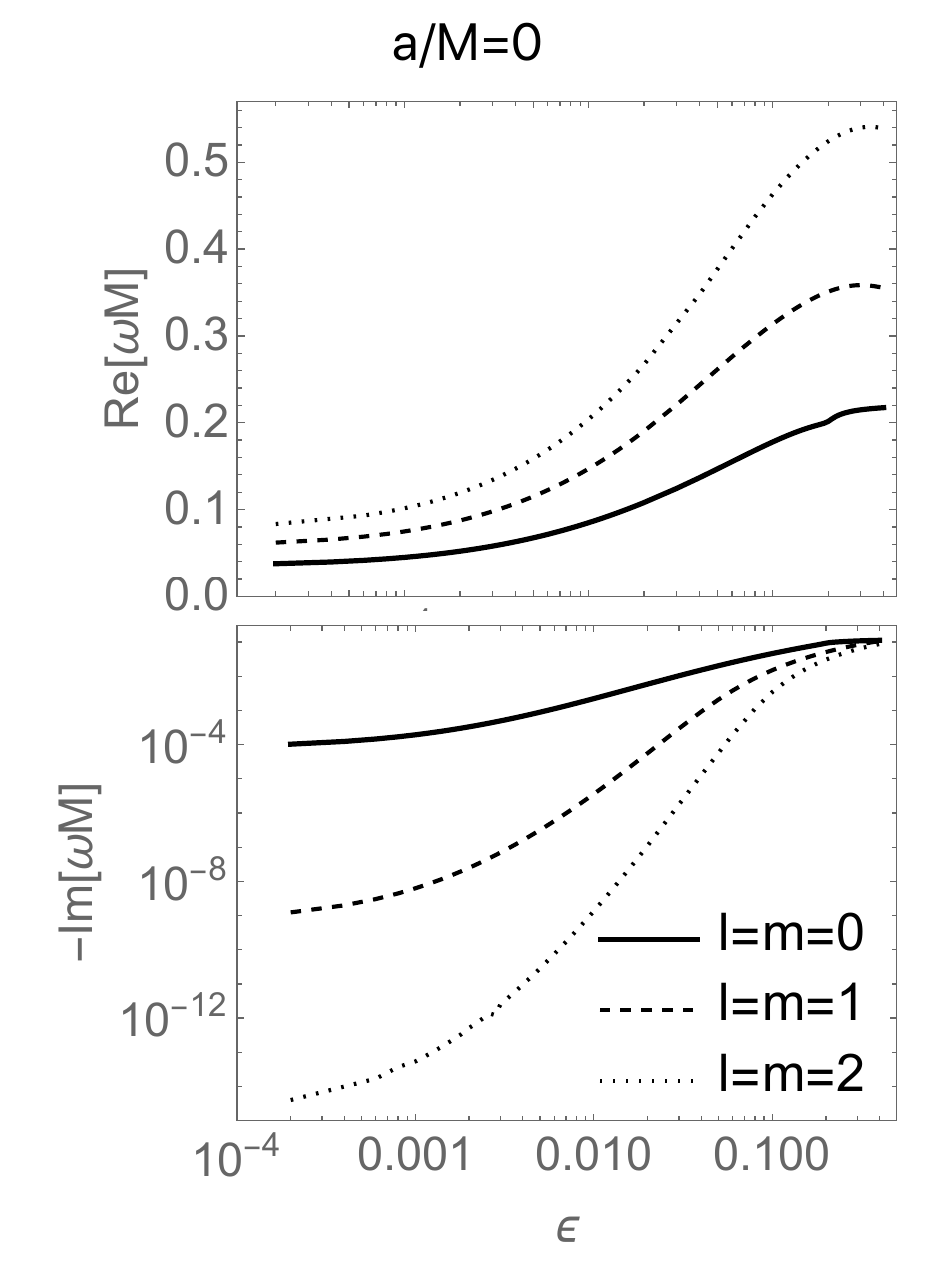}
\includegraphics[width=0.19\textwidth]{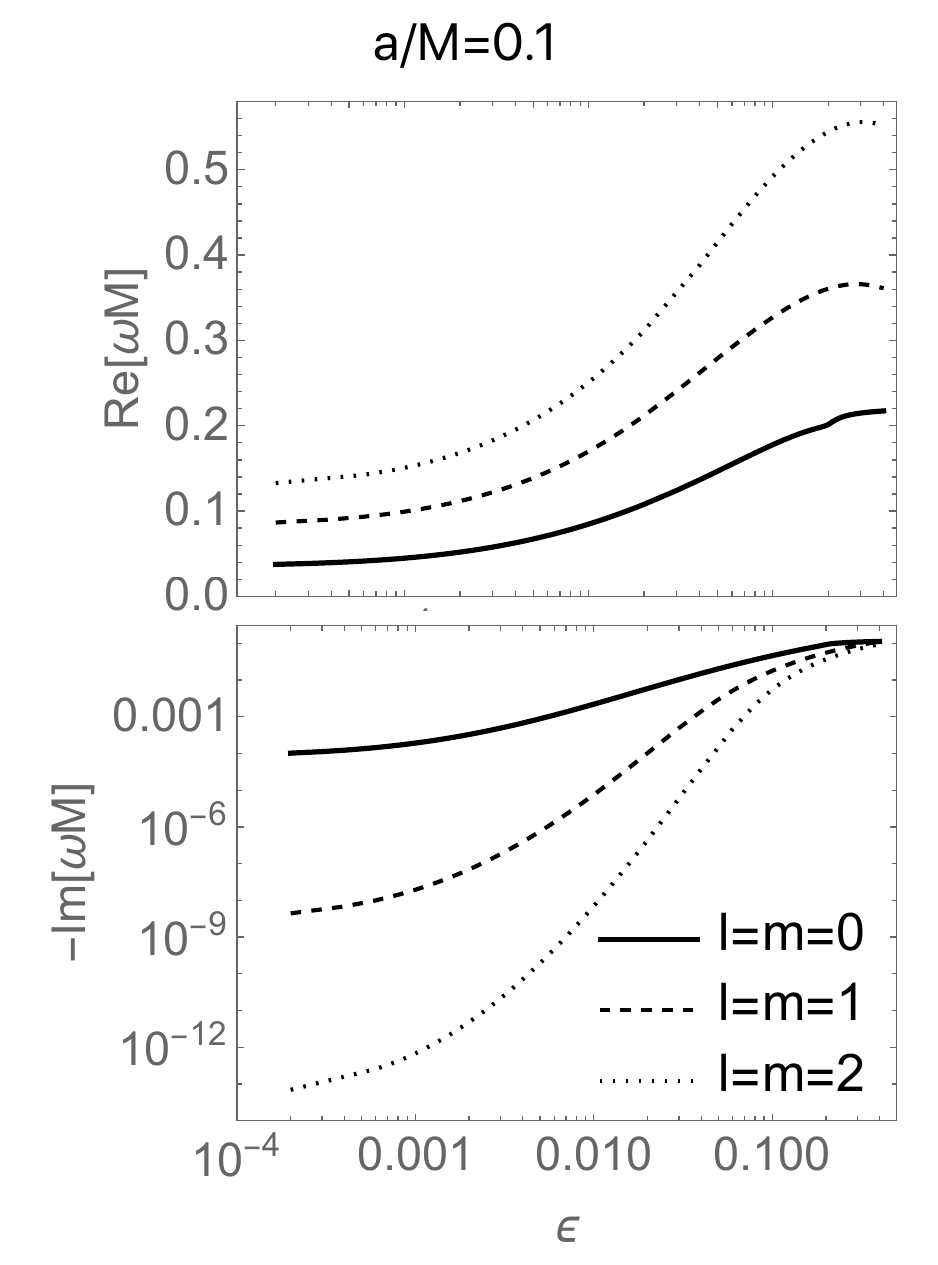}
\includegraphics[width=0.19\textwidth]{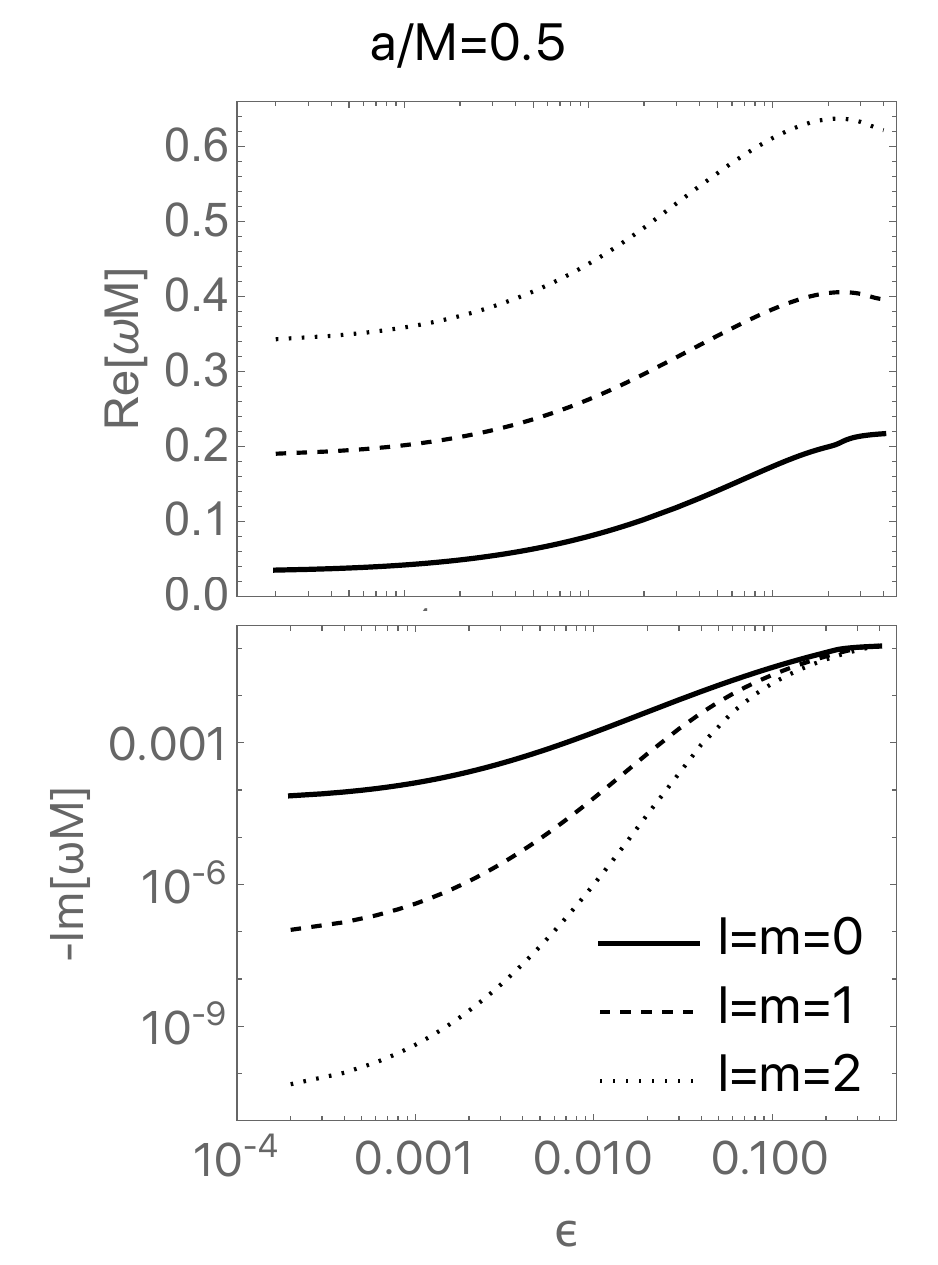}
\includegraphics[width=0.19\textwidth]{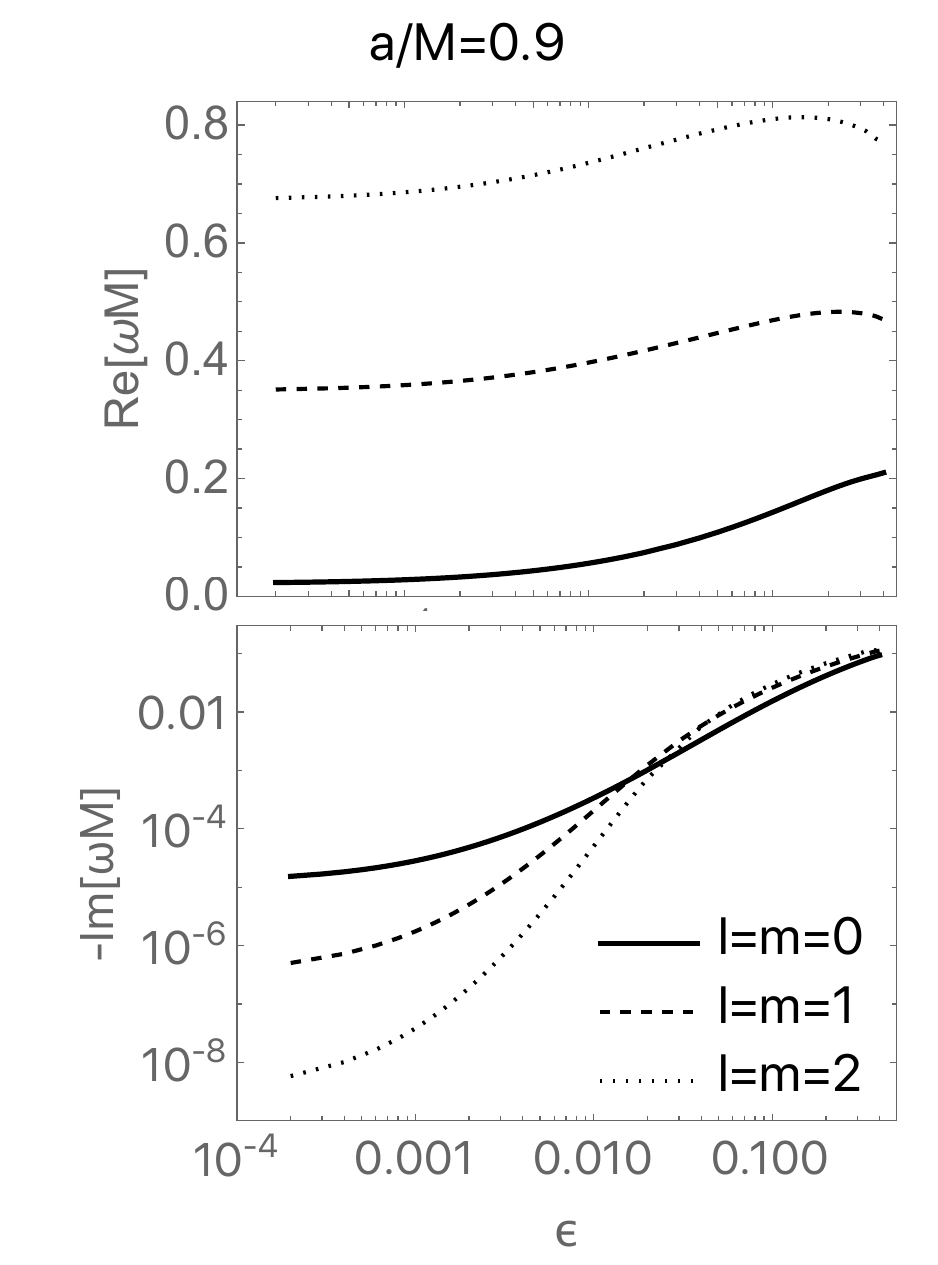}
\includegraphics[width=0.19\textwidth]{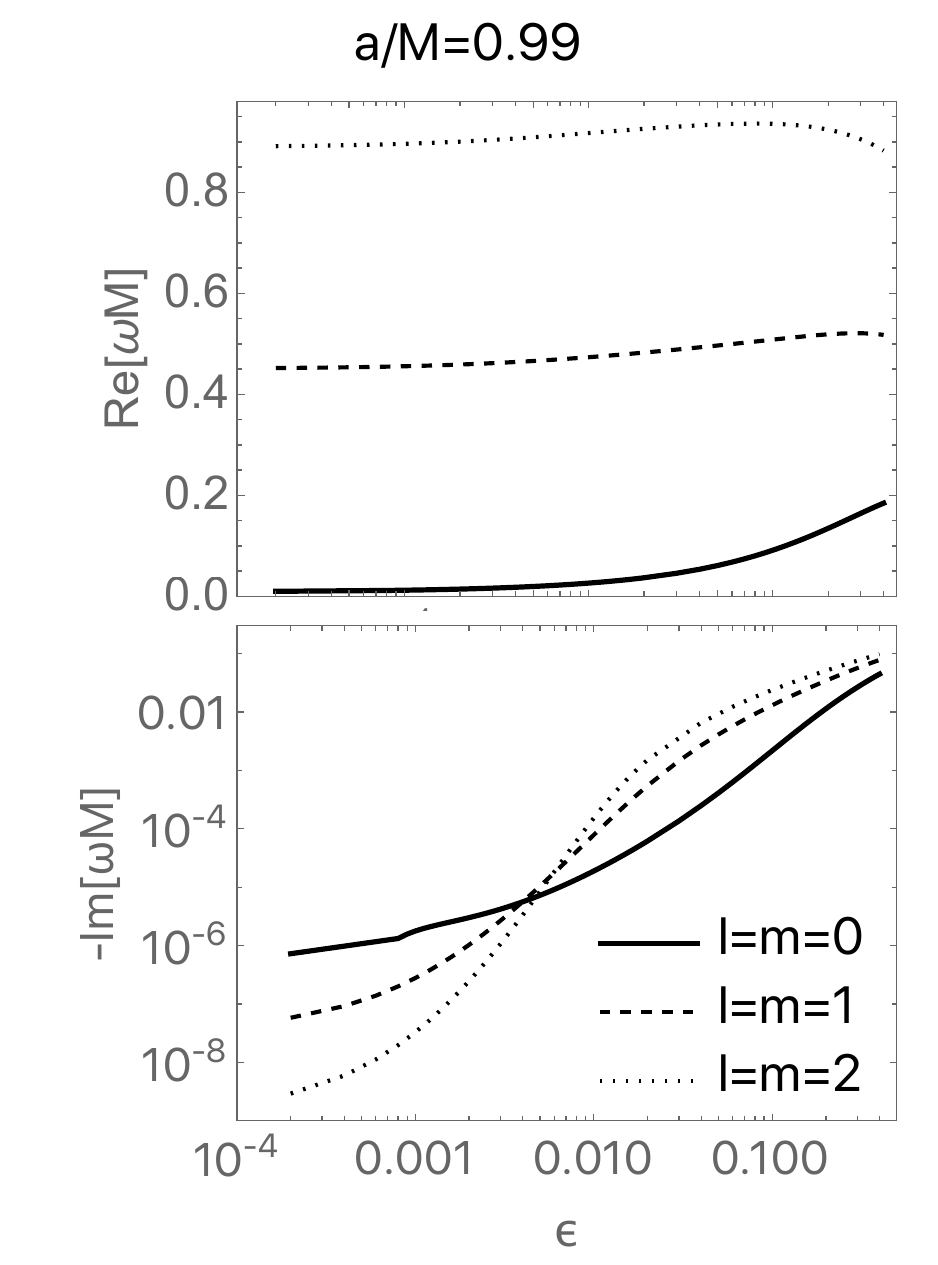}
\caption{QNMs for rotating wormholes: real (top panels) and imaginary parts (bottom panels) of the QNM frequencies have been plotted as a function of the dimensionless parameter $\epsilon:=(\ell/r_+)-1$, depicting how much the wormhole throat is shifted from the would-be black-hole horizon. We have presented the QNM frequencies for the first few $l=m$ modes, for selected values of the spin parameter.\label{fig:QNMsWH}}
\end{figure*}

As a result of this investigation, we were able to pinpoint a ``fundamental'' QNM, which we tracked under changes of the rotation parameter $a$ and regularizing parameter $\ell$ --- see \cref{fig:QNMsWH}. This mode is stable and is the least damped of a family of stable modes, which we identify as the overtones.

\begin{figure*}
\includegraphics[width=0.19\textwidth]{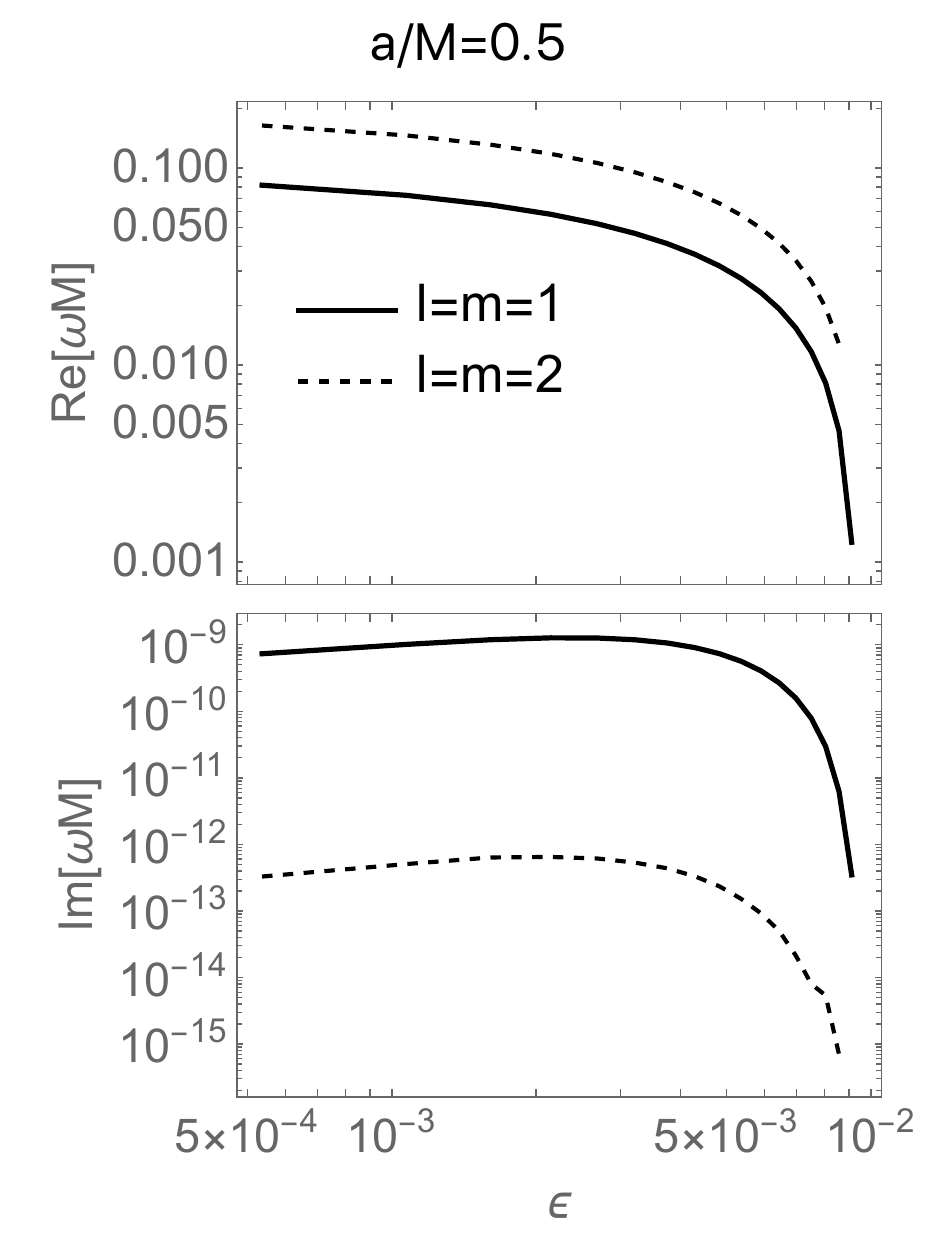}
\includegraphics[width=0.19\textwidth]{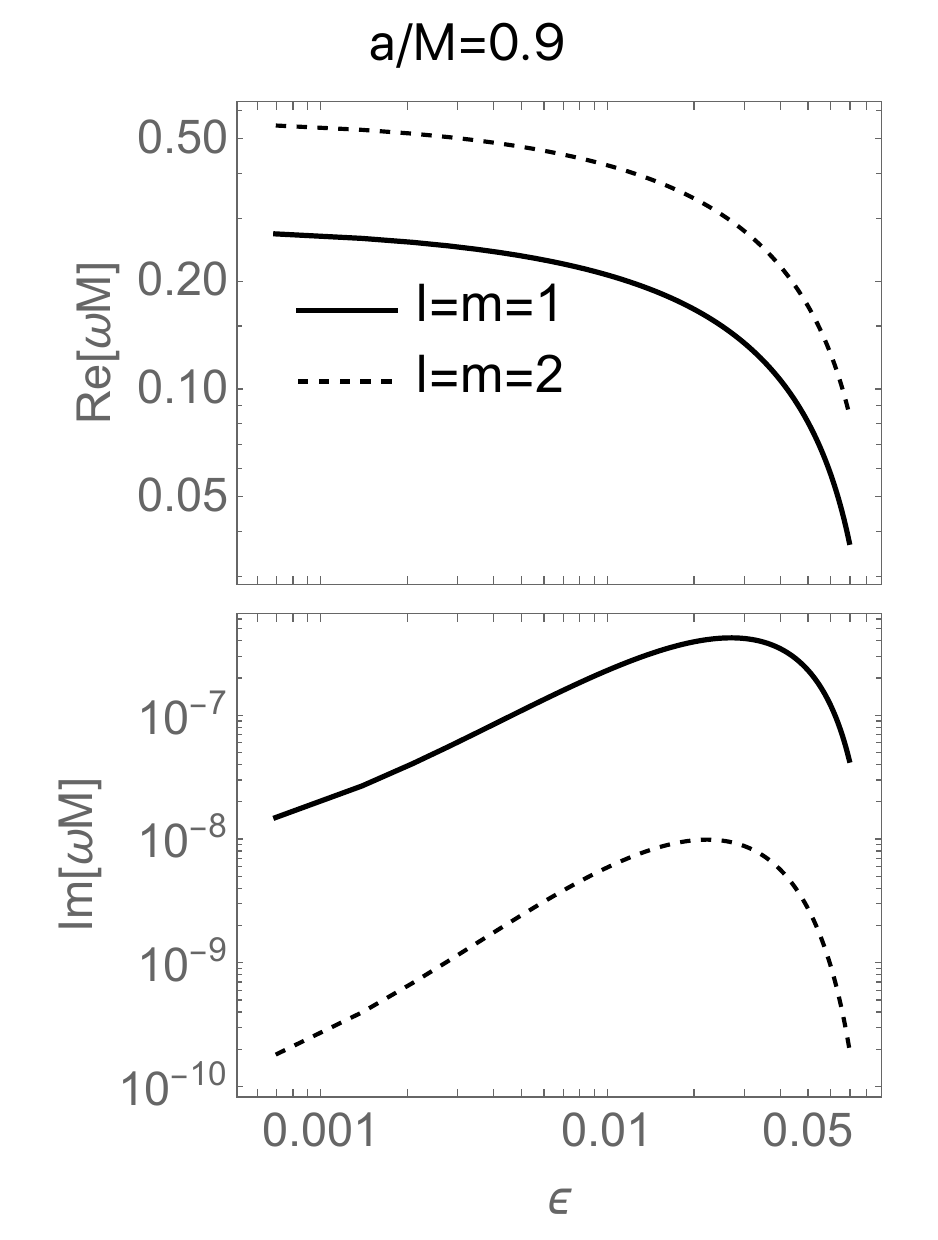}
\includegraphics[width=0.19\textwidth]{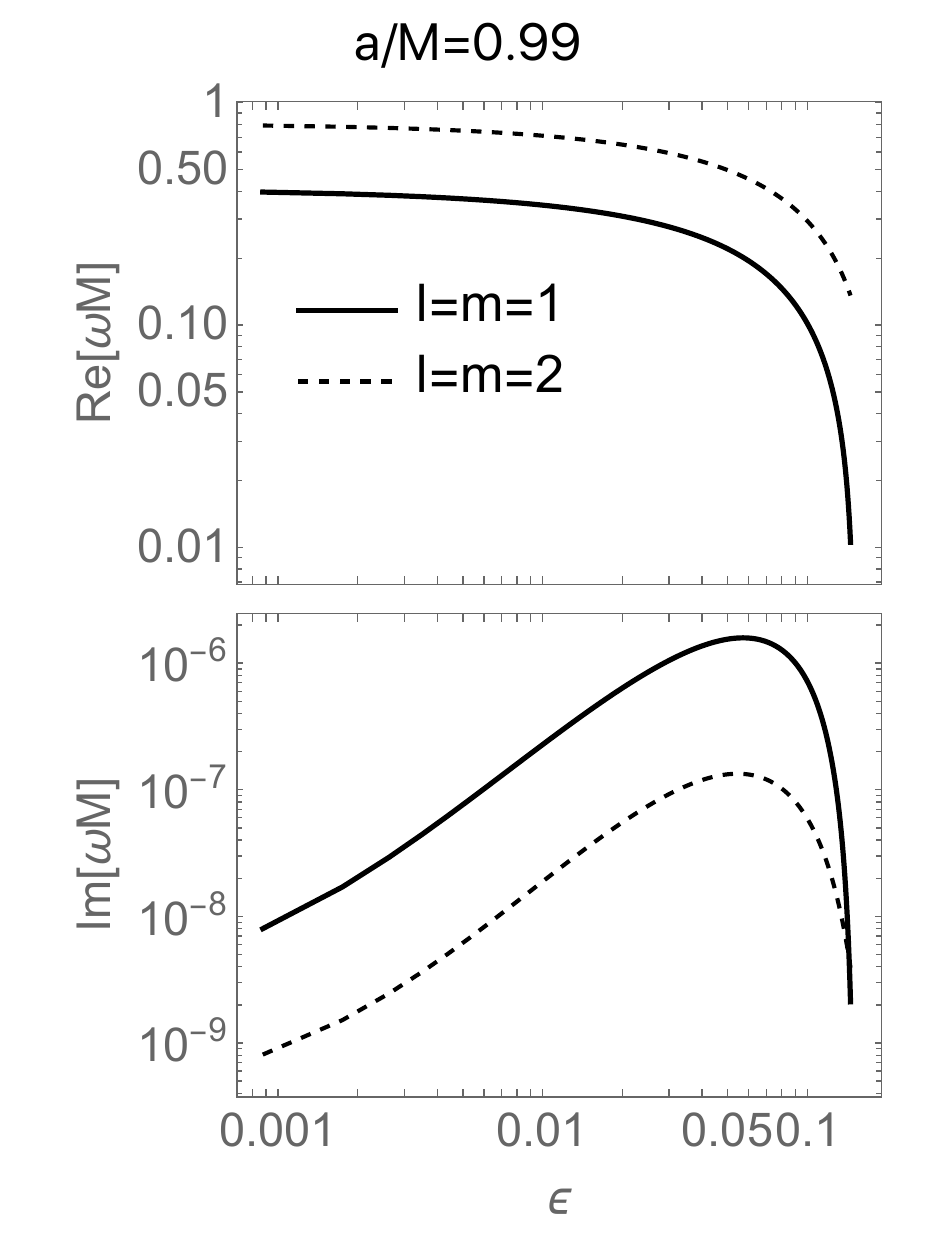}
\caption{Unstable QNMs for rotating wormholes: real (top panels) and imaginary parts (bottom panels) of the QNM frequencies have been presented as a function of the dimensionless regularizing parameter $\epsilon:=(\ell/r_+)-1$, for the first few unstable $l=m$ modes for selected values of the spin parameter. As evident, the imaginary part of the QNM frequencies are positive, signaling instability.\label{fig:WH_unstable}}
\end{figure*}

In addition to these, for high enough values of the spin parameter, and for $\Re(\omega) < m \Omega_\text{H}$, being $\Omega_\text{H}$ the would-be horizon angular velocity, a second family of QNMs appears.
All of the modes in this second family are \emph{unstable}; the imaginary parts of their QNM frequencies are very small, but \emph{positive}, and span several orders of magnitude, between approximately $10^{-6}/M$ and $10^{-15}/M$, corresponding to instability timescales in the approximate range $10\ \text{to}\ 10^{10} \left(M/M_\odot\right)\text{s}$.
For some specific cases, a few of these modes have been presented in \cref{fig:WH_unstable}.

Once again, the qualitative picture presented herein is unaffected by changes in the parameters that specify the numerical routines (the values of the numerical infinity and of the numerical throat). However, as in the null-throat wormhole case, high accuracy and precise initial guesses for the QNM frequencies are required in our numerical routine, otherwise the numerical value of the QNM frequencies found with the shooting technique could not converge. Furthermore, when changing the spacetime parameters for not-so-close-by configurations, the shooting can jump from the fundamental mode to an overtone, meaning that we had to consider a quite narrow parameters grid.
Despite these numerical difficulties, our results clearly show that there are unstable modes for the traversable wormhole configurations.

Our results on the instability timescale are compatible with those in Ref.~\cite{Cardoso:2008kj}, where Kerr-like wormholes are modeled by the Kerr metric with a mirror at finite Boyer--Lindquist radius larger than the would-be horizon.

\section{Superradiance for regular black holes\label{sec:Superradiance}}

The existence of an ergoregion, and the fact that some of its features depend on $\ell$, motivate an investigation into the phenomenon of superradiance: bosonic waves propagating on top of a Kerr black hole background can get amplified at the expense of the hole's rotational energy. It is reasonable to expect that the same will happen in a Kerr--black-bounce background, though to a different degree --- cf.~\cite{Franzin:2021kvj}. In what follows we first build an intuition on the relevant physics by analyzing the Penrose process in the vicinity of a Kerr--black-bounce regular black hole, then compute the spectrum of superradiant amplification, for massless and massive scalar fields and for different values of $\ell$.

We do not repeat the same analysis for the wormholes, as these are known to yield no supperadiant amplification according to an argument presented in Ref.~\cite{konoplya_passage_2010}. 
To understand why, think of a scattering experiment whereby a monochromatic wave, with amplitude $\mathcal{I}$, is sent from past null infinity in our universe towards the wormhole: part of the radiation will be reflected and part will be transmitted, will cross the throat and reach the future null infinity in the other universe.
Let the amplitudes of the reflected and transmitted waves --- as read off at infinity --- be $\mathcal{R}$ and $\mathcal{T}$, respectively.
As a consequence of the equation of motion, one can write the following relation
\be
-\iu \omega_\text{our} (\abs{\mathcal{I}}^2-\abs{\mathcal{R}}^2) = -\iu \omega_\text{other} \abs{\mathcal{T}}^2
\ee
(the two sides of the equation are nothing but the Wronskian, which is $r$-independent, computed at infinity in our universe, on the left, and in the other, on the right). Crucially, because of the symmetry of the spacetime the frequency of the wave at infinity in our and in the other universe coincide, $\omega_\text{our} = \omega_\text{other}$. Hence $\abs{\mathcal{R}}^2\leq \abs{\mathcal{I}}^2$, \ie\ superradiant amplification cannot happen. As already mentioned in discussing boundary conditions in \cref{subsec:bc}, alternative scenarios can be conceived; their exploration however lies beyond the scope of this work.

\subsection{The Penrose process around regular black holes\label{s:penrose}}

Classical analyses of the \emph{maximal} efficiency of the Penrose process~\cite{penrose_extraction_1971} are summarized in Ref.~\cite{chandrasekhar_mathematical_1983} --- see also~\cite{bardeen_rotating_1972, wald_energy_1974, kovetz_efficiency_1975}. In this framework, one typically considers particles on the equatorial plane and splitting at their turning points, \ie\ with vanishing radial velocities; and further notices that the most efficient extraction of energy requires both decay products to be photons. One finds
\be
\eta = \frac{E_\text{output} - E_\text{input}}{E_\text{input}} = \frac{1}{2}\left(\sqrt{1+g_{tt}}-1 \right),
\ee
where $g_{tt}$ must be evaluated at the point of splitting.
Hence, the maximal efficiency is achieved for particles splitting at the inner edge of the ergoregion and its value is governed by the magnitude of $g_{tt}$ at that point. (For an extremal Kerr black hole one finds $\eta \approx 20\%$.)

Since in our spacetime the component $g_{tt}$ is the same as in Kerr, we must conclude
\be
\eta_\text{max} = \frac{1}{2}\left( \sqrt{\frac{2M}{\tilde r}}-1\right) \qq{where} \tilde r = \max(r_+, \ell);
\ee
\ie\ the maximal efficiency of the Penrose process is completely insensitive to $\ell$ as long as this is smaller than $r_+$.

This argument, however, does not provide a complete picture of the energetics of the Penrose process. Indeed, if we aim at using it to gain insight into other processes linked to the ergoregion, we cannot limit our attention to its \emph{maximal} efficiency and the many assumptions that this brings about.
In particular, we should consider decays that take place at any point in the ergoregion, not just its inner edge, and --- crucially --- away from the turning point.

Let us stick to equatorial motion. Take a particle with mass $\mu$, energy $E$ and angular momentum along the rotation axis $L$. Its motion is effectively one-dimensional and governed by
\be\label{eq:radeom}
\frac{r^2 \Dot{r}^2}{\delta} = T
\ee
where the dot denotes differentiation with respect to an affine parameter along the geodesic and
\be\label{eq:R}
T &= \tau_1 E^2-2\tau_2 E + \tau_3\,,\\
\tau_1 &= r^4+a^2(r^2+2Mr),\\
\tau_2 &= 2aMLr,\\
\tau_3 &= L^2 a^2 - \Delta(\mu^2r^2+L^2)\,.
\ee
We may write \cref{eq:radeom} as
\be
\tau_1 E^2 -2\tau_2 E + \tilde \tau_3 = 0
\ee
with $\tilde \tau_3 = \tau_3 - r^2 \Dot{r}^2/\delta$, which has formally two roots
\be
V_\pm = \frac{\tau_2 \pm \sqrt{\tau_2^2-\tau_1 \tilde \tau_3}}{\tau_1}
= \omega L \pm \sqrt{\omega^2 L^2 - \tilde\tau_3/\tau_1}\,;
\ee
here $\omega = -g_{t\phi}/g_{\phi \phi}$ is the angular velocity of frame dragging. Actually, only the root $V_+$ is acceptable, since it must be $E > \omega L$ for the particle's momentum to be future-directed. 

Since $\delta \leq 1$, we have that $\tilde \tau_3 \leq \eval{\tilde \tau_3}_{\ell=0}$ and therefore $V_+ \geq \eval{V_+}_{\ell=0}$. Thus, particles moving in this spacetime are generically \emph{more energetic} than their counterparts in Kerr.
When in particular $E < 0$, \ie\ for Penrose's negative energy states, $\abs{E} \leq \eval{\abs{E}}_{\ell=0}$: these are ``less negative'' than their Kerr counterparts, \emph{ceteris paribus}. 

We would like to emphasize that the above analysis involving the Penrose process is a warm up exercise, while our main aim is to study superradiance. Our results demonstrate that there are certain quantities associated with the Penrose process, e.g., maximal efficiency, which are independent of the parameter $\ell$, while some others, e.g., energy extraction by a particle in radial motion with a fixed angular momentum, predict smaller values, in the same coordinate chart as Kerr. Of course, this is not conclusive and does not exhaust all possible scenarios involving the Penrose process, but is one indication towards less amount of energy being extracted from such regular black holes. This prompts us to study the superradiance of the Kerr--black-bounce spacetime in detail.

\subsection{Numerical results}

Consider first an incident massless wave with amplitude $\mathcal{I}$ coming from infinity and producing a reflected wave with amplitude $\mathcal{R}$. The asymptotic solution to \cref{radialeq} can be written as
\be\label{asymptoticR}
R \sim \mathcal{I}\,\e^{-\iu\omega r}r^{-2\iu M\omega-1} + \mathcal{R}\,\e^{\iu\omega r}r^{2\iu M\omega-1}\,.
\ee

The Kerr--black-bounce spacetime is asymptotically indistinguishable from the Kerr spacetime, hence the energy fluxes of scalar fields at infinity can be defined by the above asymptotic behavior exactly as in the Kerr spacetime~\cite{Teukolsky:1974yv}.
In particular, the ingoing and outgoing fluxes are proportional to the modulus of the amplitudes, and we can define a quantity $Z_{0,l,m}$ which gives the amplification or absorption factor for scalar waves with quantum numbers $(l,m)$ off a black hole. In this case,
\be
Z_{0,l,m} = \frac{\dd E_\text{out}}{\dd E_\text{in}} - 1 = \frac{|\mathcal{R}|^2}{|\mathcal{I}|^2} - 1\,.
\ee
In the Kerr spacetime, for massless scalar fields, this quantity can be positive only for frequencies satisfying~\cite{brito_superradiance_2020}
\be
\omega < m\Omega_\text{H}\,,
\ee
where $\Omega_\text{H}$ is the horizon angular velocity. The same reasoning can be applied to our case yielding an identical result. The angular velocity of the horizon of the regular black hole in the Kerr--black-bounce scenario, is still given by
\be
\Omega_\text{H} = \frac{a}{2Mr_+}\,,
\ee
as in Kerr. Thus, we expect the superradiant interval not to depend on $\ell$.

For generic values of the frequency, the angular and radial equation must be integrated numerically.
For each couple $(l,m)$ and value of $a\omega$ we first compute the angular eigenvalue and then we integrate the radial equation for a fixed value of $\ell$ from the horizon with ingoing boundary conditions until a sufficiently large radius.
Our numerical solution is compared to the expansion in \cref{asymptoticR} to extract the amplitudes and finally determine the amplification factor $Z_{0,l,m}$.
To increase the accuracy of these computations, we have used a higher-order expansion near the horizon and at infinity.

\begin{figure}[!thb]
\includegraphics[width=.48\textwidth]{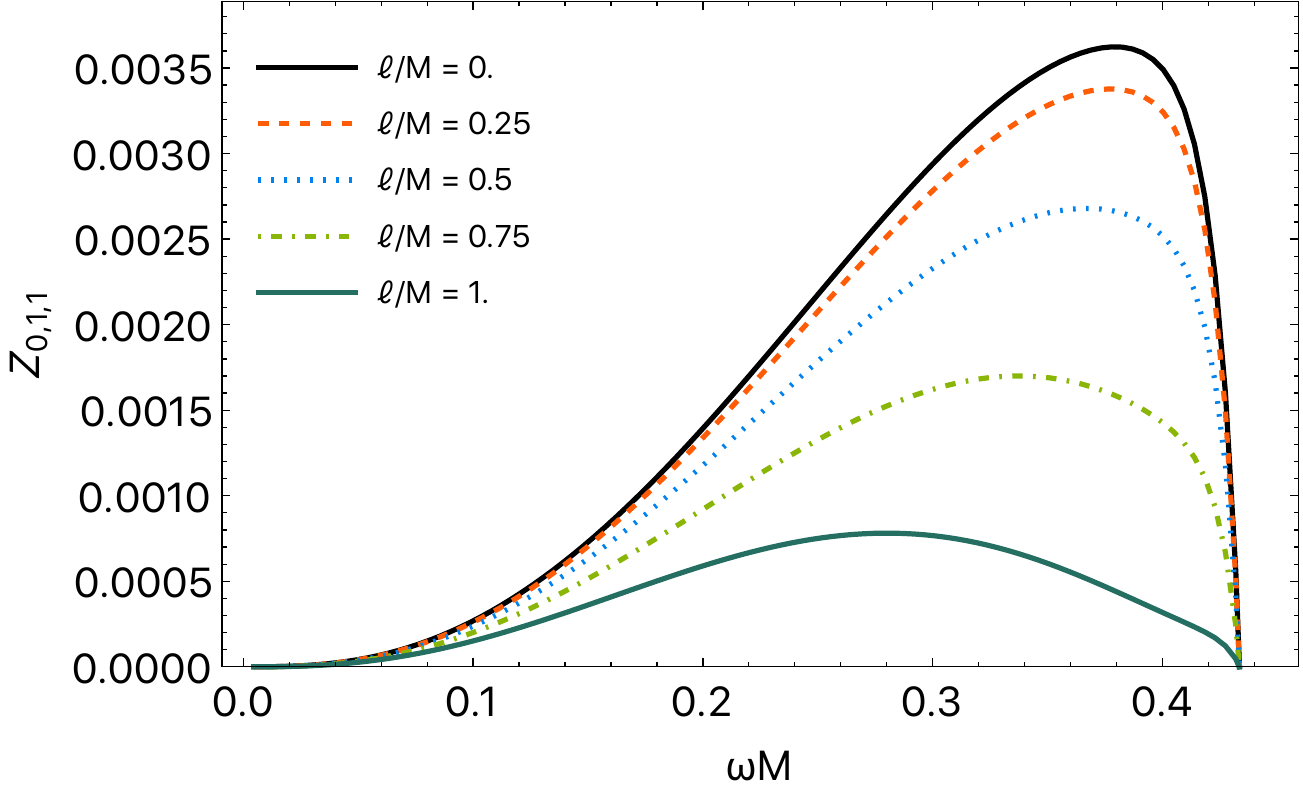}
\caption{Spectra of the amplification factor for a massless scalar with $l=m=1$ off a regular black hole with $a=0.99M$ for selected values of the regularizing parameter.\label{fig:Zmassless}}
\end{figure}

To obtain a spectrum of the amplification factor, we repeat the routine for several values of $\omega$ for different values of the black-hole parameters and the scalar field quantum numbers.
An example of our results is shown in \cref{fig:Zmassless} for an $l=m=1$ scalar wave scattered off a highly spinning black hole with $a/M=0.99$ and selected values of the regularizing parameter $\ell$.

Similarly to what happens for a Kerr black hole, the amplification factor is larger for higher values of the spin parameter and for the minimum allowed value of $l=m$, \ie\ $l=m=1$.
Modes with $m\leq0$ are not superradiant while the phenomenon is less pronounced for other values of $(l,m)$. \Cref{fig:Zmassless} confirms the general arguments in \cref{s:penrose} on the Penrose process and shows that superradiance is reduced for $\ell\neq0$ and vanishes for $\ell\to r_+$. 
Note, incidentally, that this behavior disproves the intuition according to which the spatial extent of the ergoregion determines the amount of superradiance. Indeed, as shown in \cref{app:ergo}, both the volume of the ergoregion and the area of the ergosurface actually \emph{increase} with $\ell$.
We also notice that, although the superradiant threshold frequency does not depend on $\ell$, the position and the maximum value of $Z_{0,l,m}$ do. In particular we observe a drift of position of the maximum towards smaller frequencies for larger values of $\ell/M$.
For values of the frequency larger than the superradiant threshold, the amplification factor approaches rapidly the value $-1$.

In the non-rotating limit superradiance disappears and our results agree with those of Ref.~\cite{Junior:2020lse} on the scalar absorption cross section.

\begin{figure}[!thb]
\includegraphics[width=.48\textwidth]{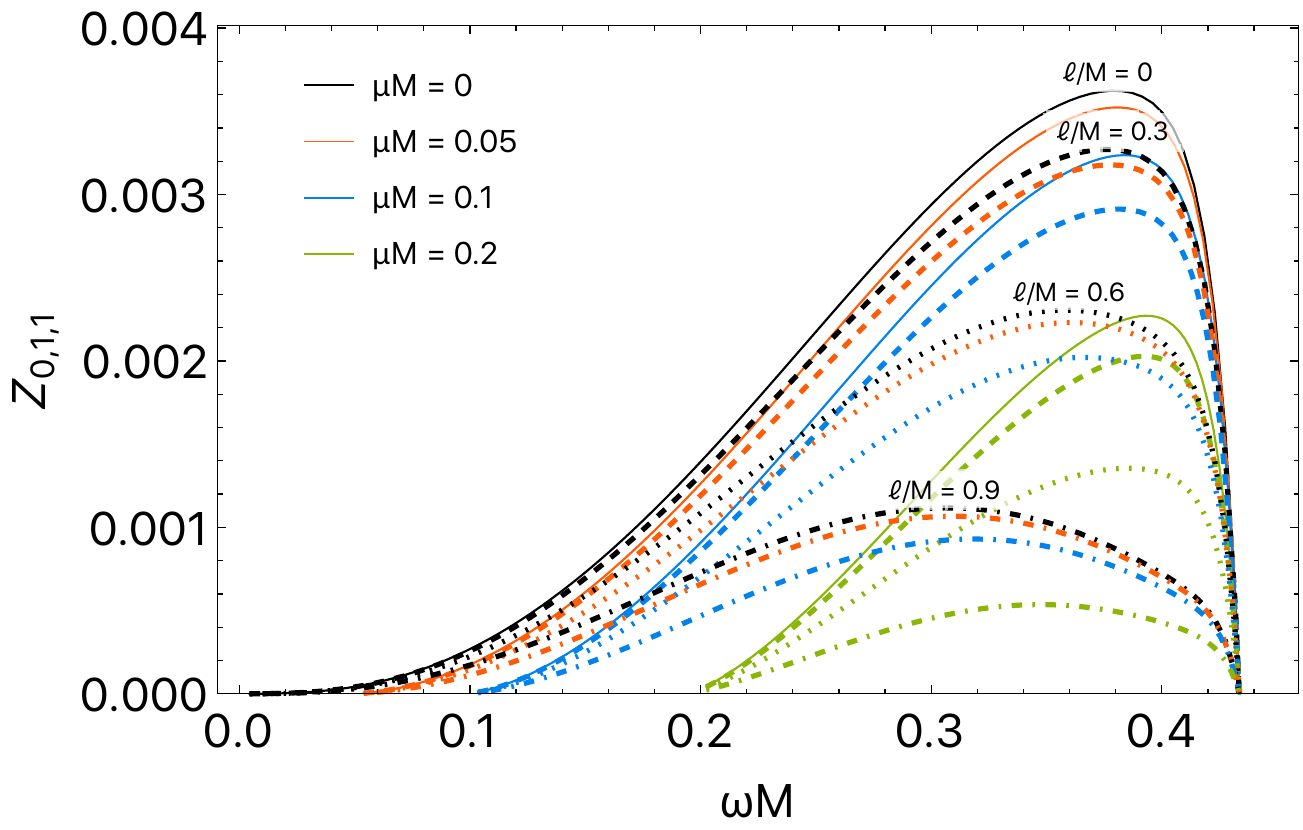}
\caption{Spectra of the amplification factor for a massive scalar with $l=m=1$ off a regular black hole with $a=0.99M$: different colours distinguish among choices of the mass parameter, while linestyles mark the values of the regularizing parameter (solid $\ell/M=0$, dashed $\ell/M=0.3$, dotted $\ell/M=0.6$, dash--dotted $\ell/M=0.9$). \label{fig:Zmassive}}
\end{figure}

These results can easily be extended to massive scalar fields. Once the appropriate boundary conditions are taken into account, the numerical procedure is identical.
In \cref{fig:Zmassive} we show spectra of the amplification factor for an $l=m=1$ scalar wave scattered off a regular black hole with $a/M=0.99$ and selected values of the regularizing parameter $\ell$ and the mass parameter $\mu$.
Massive waves can be superradiant in the frequency range $\mu<\omega<m\Omega_\text{H}$, while they are trapped near the horizon and exponentially suppressed at infinity for $\omega<\mu$.
We notice that even in this case superradiance is reduced both for larger values of $\ell/M$ and $\mu M$.
Moreover, there could be a degeneracy in the sense that the spectrum of a massive wave off a Kerr black hole might look like the spectrum of a massive (but also massless) wave off a regular black hole with the same spin.

\section{Conclusions}

In this paper we have studied and analyzed some phenomenological aspects of scalar test-field perturbations on top of the novel family of rotating black-hole mimickers proposed in Ref.~\cite{Mazza:2021rgq}.
In view of testing GR and compact objects, these geometries are appealing as they smoothly interpolate between regular black holes and traversable wormholes, depending on the value of the regularizing parameter that enters the metric.

First, we have computed the scalar QNMs.
For the regular black hole spacetime, we have used both the WKB approximation, as well as the direct integration of the scalar perturbation equation.
Our analysis demonstrates that there is a deviation of the QNM spectrum from that of Kerr black holes due to the non-zero value of the regularizing parameter~$\ell$.
Wormholes with a null throat (which coincides with the horizon) phenomenologically behave as black holes and their QNMs are in continuity with those of regular black holes. On the other hand, the presence of the throat for traversable wormholes modifies the boundary conditions, and hence the QNM spectra.
In particular, we have imposed Dirichlet boundary conditions at the throat obtaining QNM frequencies with greatly suppressed imaginary part as compared to the black-hole case and in some scenarios even positive.
This seems to indicate that rotating traversable wormholes are unstable to small perturbations. This is somewhat expected for rotating horizonless objects, albeit this instability could be tamed by relaxing the purely reflective conditions at the throat and allow for partial absorption. 

Second, we have studied the phenomenon of superradiance for both massless and massive test scalar fields around rotating regular black holes.
It turns out that both the Penrose process and superradiance are suppressed by the regularizing parameter $\ell$.
For example, in the Penrose process, the particles' energies become less negative, compared to their counterparts in Kerr, as $\ell$ gets larger (and let us stress that the faster is the black hole spin the closer has to be $\ell$ to the outer horizon in order to remove the inner horizon and the associated mass-inflation instability).
Similarly, for superradiance, the amplification of the modes depends on $\ell$: the larger $\ell/M$ the smaller the amplification factor, meaning that $\ell$ actually stabilizes the black hole against superradiant instability.
In the black-hole-to-wormhole limit $\ell\to r_+$, the amplification factor gets suppressed and it vanishes for the null-throat wormhole, while in the wormhole branch there cannot be superradiant amplification, at least as long as the wormhole is symmetric and the throat can be modeled by a purely reflective surface.
Relaxing these conditions at the throat and allowing for partial absorption, might also resolve the instability of traversable wormholes against small perturbations. 
We hope to clarify this and the above discussed open issues in future investigations.

\begin{acknowledgments}
EF, SL and JM acknowledge funding from the Italian Ministry of Education and Scientific Research (MIUR) under the grant PRIN MIUR 2017-MB8AEZ\@.
Research of SC is funded by the INSPIRE Faculty fellowship from DST, Government of India (Reg.\ No.~DST/INSPIRE/04/2018/000893) and by the
Start-Up Research Grant from SERB, DST, Government of India (Reg.\ No.~SRG/2020/000409).
\end{acknowledgments}

\appendix

\section{Properties of the ergoregion\label{app:ergo}}

Superradiance and the ensuing instability are linked to the existence of an ergoregion, \ie\ a portion of the spacetime in which the Killing vector associated to time translations --- which is timelike at spatial infinity --- becomes spacelike. With this appendix, we aim at spelling out some of its relevant details in Kerr--black-bounce spacetimes. A quick inspection of the metric in \cref{background} allows to identify the ergoregion with the locus of points for which $\Sigma -2M r \leq 0$. Equality is met at
\be
r = r^\pm_\text{erg}(\theta):= M \pm \sqrt{M^2 -a^2 \cos^2 \theta}\,.
\ee
When $a>M$ --- a case we never consider in this article --- there are no horizons and the curves $r^\pm_\text{erg}(\theta)$, along with the throat $r=\ell$, mark the boundary of the ergoregion; note that for $\ell > 2M$ no ergoregion exists. When instead $a \leq M$, the ergoregion stretches between $r^+_\text{erg}(\theta)$ and the horizon, if there is one, or the wormhole throat. When $\ell > r_+$, in particular, the ergosurface does not extend to the poles but is limited to polar angles $\theta \in \left[\theta_*,\ \pi-\theta_* \right]$, with
\be
\theta_* := \arccos \left(\frac{\sqrt{\ell(2M-\ell)}}{a} \right);
\ee
\ie\ it is a solid of revolution whose section is shaped as a crescent and whose axis coincides with the axis of symmetry of the spacetime.
At any given time, the area of the ergosurface is given by the integral~\cite{pelavas_properties_2001}
\be
A_\text{erg} = \int\dd\theta\,\dd\phi\,\sqrt{S} 
\ee
where $\phi \in \left[0, 2\pi\right]$, $\theta \in \left[0, \pi \right]$ or $\theta \in \left[ \theta_*, \pi-\theta_* \right]$ when $\ell > r_+$ and $S$ is the determinant of the two-dimensional induced metric. Specifically, we have 
\be
S &= \left[g_{rr} \left(\frac{\dd r_\text{erg}}{\dd\theta} \right)^2+g_{\theta\theta}\right]g_{\phi \phi}\0\\
&= \Sigma \left[1+\frac{a^2 \cos^2 \theta}{\delta (r_\text{erg}^2-M^2)} \right]2\sin^2 \theta \left( M r_\text{erg}+a^2\sin^2 \theta\right),
\ee
which should be evaluated at $r=r_\text{erg}$. 

Since $\delta\leq 1$, as long as $\ell < r_+$, we expect $A_\text{erg} \geq \eval{A_\text{erg} }_{\ell=0}$, \ie\ that the area be larger than its Kerr analogue; for $\ell>r_+$, instead, $A_\text{erg}$ is a continuously decreasing function of $\ell$ that reaches zero for $\ell=2M$. Note, incidentally, that surfaces of constant $r$, such as the horizon, have the same area in our spacetime as they have in Kerr: the $\ell$-dependence comes in as soon as different radii are spanned.

The volume of a constant-$t$ slice of the ergoregion is given by
\be\label{Verg}
V_\text{erg}=\int\dd{}r\,\dd\theta\,\dd\phi\,\sqrt{h} 
\ee
where $r\in \left[\max(r_+, \ell), r_\text{erg}\right]$, $\theta \in \left[0, \pi\right]$ when $\ell \leq r_+$ or $\theta \in \left[\theta_*, \pi-\theta_* \right]$ otherwise, and $\phi \in \left[0, 2\pi\right]$. We have
\be
h = g_{rr}g_{\theta \theta}g_{\phi \phi} = \frac{\Sigma A \sin^2 \theta}{\Delta \delta}\,.
\ee
The integrand in \cref{Verg} has poles at $r=r_+$ and $r=\ell$, \ie\ along the inner edge of the ergoregion. The integral itself is usually convergent, unless $\ell=r_+$: in this case the two poles coincide and the integral diverges logarithmically. Something analogous happens for extremal Kerr black holes, see~\cite{Pani:2010jz}. In any case, since $\delta \leq 1$, the volume of the ergoregion will be larger than that of the corresponding Kerr as long as $\ell< r_+$; for larger values of $\ell$, instead, the volume will strictly decrease and reach zero for $\ell = 2M$.

The fact that both the area of the ergosurface and the volume of the ergoregion increase with increasing $\ell$, while superradiant amplification gets tamed, disproves the intuitive notion that a larger ergoregion entails ``more superradiance''. A better understanding of the physics of this phenomenon is provided by the analysis of the Penrose process in the equatorial plane given in \cref{s:penrose}.

\section{Singularities of the radial equation\label{appendix}}

In this appendix, we elucidate some subtleties concerning the behavior of the solution to the radial \cref{radialeq} close to its singular points. First of all, write \cref{radialeq} in canonical form:
\be
R'' + \alpha(r) R' + \beta(r) R = 0\,,
\ee
where
\be
\alpha(r) &= \frac{\left(\sqrt{\delta} \Delta \right)'}{\sqrt{\delta} \Delta}\,,\\
\beta(r) &= \frac{1}{\delta \Delta^2}\left(\frac{\left[\left(r^2+a^2\right)\omega - a m\right]^2}{\Delta} - \lambda - \mu^2 r^2\right).
\ee
Note that 
\be
\alpha(r) =
%\frac{1}{2} \frac{\delta'}{\delta}+\frac{\Delta'}{\Delta} =
\frac{1}{2}\left[\frac{1}{r+\ell} + \frac{1}{r-\ell} \right]-\frac{1}{r} + \frac{1}{r-r_+}+\frac{1}{r-r_-}\,.
\ee
The poles of the coefficients $\alpha$ and $\beta$ are singular points for the differential equation. Following standard terminology~\cite{bender_advanced_1999}, we call \emph{irregular} those singular points where $\alpha(r)$ or $\beta(r)$ have a pole of order higher than one or two, respectively, and \emph{regular} the singular points where the divergences of $\alpha(r)$ and $\beta(r)$ are less severe. According to this convention, we find
\begin{itemize}
\item regular singular points at $r = r_+,\ r_-,\ 0,\ +\ell$ (and $-\ell$, technically) when $\ell \neq r_\pm$, and
\item an irregular singular point at $r=\infty$.
\end{itemize}
As $\ell\to 0$, the three poles of $\alpha$ located at $r=\pm \ell, \ 0$ exactly cancel each other out and the two poles $r=\pm \ell$ in $\beta$ also disappear; the resulting equation (second-order ODE with two regular and one irregular singular points) is of the confluent Heun type.
The ``confluent'' case in which $\ell=r_\pm$ is particularly nasty, as two regular singular points merge into an irregular singular point.

Using the throat-penetrating coordinate $r'$ instead of the Johannsen coordinate $r$ does not change the picture: $\delta$ disappears from the equation but $\Delta$ is not a polynomial of degree two and its zeroes have a more complicated structure.

In the vicinity of a regular singular point $r_0$, the equation admits a (possibly divergent) power-series solution of the form (Frobenius' method)
\be
R(r) = (r-r_0)^s \sum_{n\in \mathbb{N}} a_n(r-r_0)^n
\ee
with $s$ satisfying the \emph{indicial equation}
\be
s(s-1) +\alpha_0 s+\beta_0=0\,;
\ee
here
\be
\alpha_0 &= \lim_{r\to r_0}(r-r_0) \alpha(r) \qq{and} \beta_0 = \lim_{r\to r_0}(r-r_0)^2 \beta(r)\,.
\ee
Close to $r=r_+$, we have $s= \pm \iu \frac{a m -2M\omega r_+}{(r_+-r_-)\gamma} $, hence \cref{leading_bh}. Similarly, close to $r= \ell$, we find $s=0, \ 1/2$, although $s=0$ must be excluded since it does not give rise to a solution.

Close to an irregular singular point, one can construct a generalization of Frobenius' series.
The solution will consist of an exponential prefactor, encoding the leading divergent behavior, and a power series in the variable $(r-r_0)^c$, with $c$ some number.
Proceeding in this way, one can recover the standard result of \cref{Rasympmass}.
More interestingly, in the particular case $\ell=r_+$, close to $r=\ell$ the solution turns out to be
\be
R(r) = \exp(\pm \iu \frac{am-2M\omega r_+}{r_+ - r_-} \sqrt{\frac{2\ell}{r-\ell}}) \sum_{n\in \mathbb{N}} a_n (r-\ell)^{n/2}\,,
\ee
(hence, in particular, $c=1/2$). Such behavior renders the numerical integration of the null-throat wormhole case particularly difficult.

\bibliography{refs}

\begin{thebibliography}{76}%
\makeatletter
\providecommand \@ifxundefined [1]{%
 \@ifx{#1\undefined}
}%
\providecommand \@ifnum [1]{%
 \ifnum #1\expandafter \@firstoftwo
 \else \expandafter \@secondoftwo
 \fi
}%
\providecommand \@ifx [1]{%
 \ifx #1\expandafter \@firstoftwo
 \else \expandafter \@secondoftwo
 \fi
}%
\providecommand \natexlab [1]{#1}%
\providecommand \enquote  [1]{``#1''}%
\providecommand \bibnamefont  [1]{#1}%
\providecommand \bibfnamefont [1]{#1}%
\providecommand \citenamefont [1]{#1}%
\providecommand \href@noop [0]{\@secondoftwo}%
\providecommand \href [0]{\begingroup \@sanitize@url \@href}%
\providecommand \@href[1]{\@@startlink{#1}\@@href}%
\providecommand \@@href[1]{\endgroup#1\@@endlink}%
\providecommand \@sanitize@url [0]{\catcode `\\12\catcode `\$12\catcode
  `\&12\catcode `\#12\catcode `\^12\catcode `\_12\catcode `\%12\relax}%
\providecommand \@@startlink[1]{}%
\providecommand \@@endlink[0]{}%
\providecommand \url  [0]{\begingroup\@sanitize@url \@url }%
\providecommand \@url [1]{\endgroup\@href {#1}{\urlprefix }}%
\providecommand \urlprefix  [0]{URL }%
\providecommand \Eprint [0]{\href }%
\providecommand \doibase [0]{https://doi.org/}%
\providecommand \selectlanguage [0]{\@gobble}%
\providecommand \bibinfo  [0]{\@secondoftwo}%
\providecommand \bibfield  [0]{\@secondoftwo}%
\providecommand \translation [1]{[#1]}%
\providecommand \BibitemOpen [0]{}%
\providecommand \bibitemStop [0]{}%
\providecommand \bibitemNoStop [0]{.\EOS\space}%
\providecommand \EOS [0]{\spacefactor3000\relax}%
\providecommand \BibitemShut  [1]{\csname bibitem#1\endcsname}%
\let\auto@bib@innerbib\@empty
%</preamble>
\bibitem [{\citenamefont {Abbott}\ \emph {et~al.}(2016)\citenamefont {Abbott}
  \emph {et~al.}}]{LIGOScientific:2016aoc}%
  \BibitemOpen
  \bibfield  {author} {\bibinfo {author} {\bibfnamefont {B.~P.}\ \bibnamefont
  {Abbott}} \emph {et~al.} (\bibinfo {collaboration} {LIGO Scientific,
  Virgo}),\ }\bibfield  {title} {\bibinfo {title} {{Observation of
  Gravitational Waves from a Binary Black Hole Merger}},\ }\href
  {https://doi.org/10.1103/PhysRevLett.116.061102} {\bibfield  {journal}
  {\bibinfo  {journal} {Phys. Rev. Lett.}\ }\textbf {\bibinfo {volume} {116}},\
  \bibinfo {pages} {061102} (\bibinfo {year} {2016})},\ \Eprint
  {https://arxiv.org/abs/1602.03837} {arXiv:1602.03837 [gr-qc]} \BibitemShut
  {NoStop}%
\bibitem [{\citenamefont {Abbott}\ \emph {et~al.}(2019)\citenamefont {Abbott}
  \emph {et~al.}}]{LIGOScientific:2018mvr}%
  \BibitemOpen
  \bibfield  {author} {\bibinfo {author} {\bibfnamefont {B.~P.}\ \bibnamefont
  {Abbott}} \emph {et~al.} (\bibinfo {collaboration} {LIGO Scientific,
  Virgo}),\ }\bibfield  {title} {\bibinfo {title} {{GWTC-1: A
  Gravitational-Wave Transient Catalog of Compact Binary Mergers Observed by
  LIGO and Virgo during the First and Second Observing Runs}},\ }\href
  {https://doi.org/10.1103/PhysRevX.9.031040} {\bibfield  {journal} {\bibinfo
  {journal} {Phys. Rev. X}\ }\textbf {\bibinfo {volume} {9}},\ \bibinfo {pages}
  {031040} (\bibinfo {year} {2019})},\ \Eprint
  {https://arxiv.org/abs/1811.12907} {arXiv:1811.12907 [astro-ph.HE]}
  \BibitemShut {NoStop}%
\bibitem [{\citenamefont {Abbott}\ \emph
  {et~al.}(2021{\natexlab{a}})\citenamefont {Abbott} \emph
  {et~al.}}]{LIGOScientific:2020ibl}%
  \BibitemOpen
  \bibfield  {author} {\bibinfo {author} {\bibfnamefont {R.}~\bibnamefont
  {Abbott}} \emph {et~al.} (\bibinfo {collaboration} {LIGO Scientific,
  Virgo}),\ }\bibfield  {title} {\bibinfo {title} {{GWTC-2: Compact Binary
  Coalescences Observed by LIGO and Virgo During the First Half of the Third
  Observing Run}},\ }\href {https://doi.org/10.1103/PhysRevX.11.021053}
  {\bibfield  {journal} {\bibinfo  {journal} {Phys. Rev. X}\ }\textbf {\bibinfo
  {volume} {11}},\ \bibinfo {pages} {021053} (\bibinfo {year}
  {2021}{\natexlab{a}})},\ \Eprint {https://arxiv.org/abs/2010.14527}
  {arXiv:2010.14527 [gr-qc]} \BibitemShut {NoStop}%
\bibitem [{\citenamefont {Abbott}\ \emph
  {et~al.}(2021{\natexlab{b}})\citenamefont {Abbott} \emph
  {et~al.}}]{LIGOScientific:2021djp}%
  \BibitemOpen
  \bibfield  {author} {\bibinfo {author} {\bibfnamefont {R.}~\bibnamefont
  {Abbott}} \emph {et~al.} (\bibinfo {collaboration} {LIGO Scientific, VIRGO,
  KAGRA}),\ }\href@noop {} {\bibinfo {title} {{GWTC-3: Compact Binary
  Coalescences Observed by LIGO and Virgo During the Second Part of the Third
  Observing Run}}} (\bibinfo {year} {2021}{\natexlab{b}}),\ \Eprint
  {https://arxiv.org/abs/2111.03606} {arXiv:2111.03606 [gr-qc]} \BibitemShut
  {NoStop}%
\bibitem [{\citenamefont {Akiyama}\ \emph {et~al.}(2019)\citenamefont {Akiyama}
  \emph {et~al.}}]{EventHorizonTelescope:2019dse}%
  \BibitemOpen
  \bibfield  {author} {\bibinfo {author} {\bibfnamefont {K.}~\bibnamefont
  {Akiyama}} \emph {et~al.} (\bibinfo {collaboration} {Event Horizon
  Telescope}),\ }\bibfield  {title} {\bibinfo {title} {{First M87 Event Horizon
  Telescope Results. I. The Shadow of the Supermassive Black Hole}},\ }\href
  {https://doi.org/10.3847/2041-8213/ab0ec7} {\bibfield  {journal} {\bibinfo
  {journal} {Astrophys. J. Lett.}\ }\textbf {\bibinfo {volume} {875}},\
  \bibinfo {pages} {L1} (\bibinfo {year} {2019})},\ \Eprint
  {https://arxiv.org/abs/1906.11238} {arXiv:1906.11238 [astro-ph.GA]}
  \BibitemShut {NoStop}%
\bibitem [{\citenamefont {Damour}\ and\ \citenamefont
  {Solodukhin}(2007)}]{Damour:2007ap}%
  \BibitemOpen
  \bibfield  {author} {\bibinfo {author} {\bibfnamefont {T.}~\bibnamefont
  {Damour}}\ and\ \bibinfo {author} {\bibfnamefont {S.~N.}\ \bibnamefont
  {Solodukhin}},\ }\bibfield  {title} {\bibinfo {title} {{Wormholes as black
  hole foils}},\ }\href {https://doi.org/10.1103/PhysRevD.76.024016} {\bibfield
   {journal} {\bibinfo  {journal} {Phys. Rev. D}\ }\textbf {\bibinfo {volume}
  {76}},\ \bibinfo {pages} {024016} (\bibinfo {year} {2007})},\ \Eprint
  {https://arxiv.org/abs/0704.2667} {arXiv:0704.2667 [gr-qc]} \BibitemShut
  {NoStop}%
\bibitem [{\citenamefont {Cardoso}\ \emph
  {et~al.}(2016{\natexlab{a}})\citenamefont {Cardoso}, \citenamefont
  {Franzin},\ and\ \citenamefont {Pani}}]{Cardoso:2016rao}%
  \BibitemOpen
  \bibfield  {author} {\bibinfo {author} {\bibfnamefont {V.}~\bibnamefont
  {Cardoso}}, \bibinfo {author} {\bibfnamefont {E.}~\bibnamefont {Franzin}},\
  and\ \bibinfo {author} {\bibfnamefont {P.}~\bibnamefont {Pani}},\ }\bibfield
  {title} {\bibinfo {title} {{Is the gravitational-wave ringdown a probe of the
  event horizon?}},\ }\href {https://doi.org/10.1103/PhysRevLett.116.171101}
  {\bibfield  {journal} {\bibinfo  {journal} {Phys. Rev. Lett.}\ }\textbf
  {\bibinfo {volume} {116}},\ \bibinfo {pages} {171101} (\bibinfo {year}
  {2016}{\natexlab{a}})},\ \bibinfo {note} {[Erratum: Phys. Rev. Lett.
  {\bf117}, 089902(E) (2016)]},\ \Eprint {https://arxiv.org/abs/1602.07309}
  {arXiv:1602.07309 [gr-qc]} \BibitemShut {NoStop}%
\bibitem [{\citenamefont {Carballo-Rubio}\ \emph
  {et~al.}(2020{\natexlab{a}})\citenamefont {Carballo-Rubio}, \citenamefont
  {Di~Filippo}, \citenamefont {Liberati},\ and\ \citenamefont
  {Visser}}]{Carballo-Rubio:2019nel}%
  \BibitemOpen
  \bibfield  {author} {\bibinfo {author} {\bibfnamefont {R.}~\bibnamefont
  {Carballo-Rubio}}, \bibinfo {author} {\bibfnamefont {F.}~\bibnamefont
  {Di~Filippo}}, \bibinfo {author} {\bibfnamefont {S.}~\bibnamefont
  {Liberati}},\ and\ \bibinfo {author} {\bibfnamefont {M.}~\bibnamefont
  {Visser}},\ }\bibfield  {title} {\bibinfo {title} {{Opening the Pandora's box
  at the core of black holes}},\ }\href
  {https://doi.org/10.1088/1361-6382/ab8141} {\bibfield  {journal} {\bibinfo
  {journal} {Class. Quantum Grav.}\ }\textbf {\bibinfo {volume} {37}},\
  \bibinfo {pages} {145005} (\bibinfo {year} {2020}{\natexlab{a}})},\ \Eprint
  {https://arxiv.org/abs/1908.03261} {arXiv:1908.03261 [gr-qc]} \BibitemShut
  {NoStop}%
\bibitem [{\citenamefont {Carballo-Rubio}\ \emph
  {et~al.}(2020{\natexlab{b}})\citenamefont {Carballo-Rubio}, \citenamefont
  {Di~Filippo}, \citenamefont {Liberati},\ and\ \citenamefont
  {Visser}}]{Carballo-Rubio:2019fnb}%
  \BibitemOpen
  \bibfield  {author} {\bibinfo {author} {\bibfnamefont {R.}~\bibnamefont
  {Carballo-Rubio}}, \bibinfo {author} {\bibfnamefont {F.}~\bibnamefont
  {Di~Filippo}}, \bibinfo {author} {\bibfnamefont {S.}~\bibnamefont
  {Liberati}},\ and\ \bibinfo {author} {\bibfnamefont {M.}~\bibnamefont
  {Visser}},\ }\bibfield  {title} {\bibinfo {title} {{Geodesically complete
  black holes}},\ }\href {https://doi.org/10.1103/PhysRevD.101.084047}
  {\bibfield  {journal} {\bibinfo  {journal} {Phys. Rev. D}\ }\textbf {\bibinfo
  {volume} {101}},\ \bibinfo {pages} {084047} (\bibinfo {year}
  {2020}{\natexlab{b}})},\ \Eprint {https://arxiv.org/abs/1911.11200}
  {arXiv:1911.11200 [gr-qc]} \BibitemShut {NoStop}%
\bibitem [{\citenamefont {Simpson}\ and\ \citenamefont
  {Visser}(2019)}]{Simpson:2018tsi}%
  \BibitemOpen
  \bibfield  {author} {\bibinfo {author} {\bibfnamefont {A.}~\bibnamefont
  {Simpson}}\ and\ \bibinfo {author} {\bibfnamefont {M.}~\bibnamefont
  {Visser}},\ }\bibfield  {title} {\bibinfo {title} {{Black-bounce to
  traversable wormhole}},\ }\href
  {https://doi.org/10.1088/1475-7516/2019/02/042} {\bibfield  {journal}
  {\bibinfo  {journal} {J. Cosmol. Astropart. Phys.}\ }\textbf {\bibinfo
  {volume} {\normalfont02}}\bibfield  {number} {\bibinfo  {number} { (2019)},\
  \bibinfo {pages} {042}},\ }\Eprint {https://arxiv.org/abs/1812.07114}
  {arXiv:1812.07114 [gr-qc]} \BibitemShut {NoStop}%
\bibitem [{\citenamefont {Banerjee}\ \emph {et~al.}(2020)\citenamefont
  {Banerjee}, \citenamefont {Chakraborty},\ and\ \citenamefont
  {SenGupta}}]{Banerjee:2019nnj}%
  \BibitemOpen
  \bibfield  {author} {\bibinfo {author} {\bibfnamefont {I.}~\bibnamefont
  {Banerjee}}, \bibinfo {author} {\bibfnamefont {S.}~\bibnamefont
  {Chakraborty}},\ and\ \bibinfo {author} {\bibfnamefont {S.}~\bibnamefont
  {SenGupta}},\ }\bibfield  {title} {\bibinfo {title} {{Silhouette of M87*: A
  New Window to Peek into the World of Hidden Dimensions}},\ }\href
  {https://doi.org/10.1103/PhysRevD.101.041301} {\bibfield  {journal} {\bibinfo
   {journal} {Phys. Rev. D}\ }\textbf {\bibinfo {volume} {101}},\ \bibinfo
  {pages} {041301(R)} (\bibinfo {year} {2020})},\ \Eprint
  {https://arxiv.org/abs/1909.09385} {arXiv:1909.09385 [gr-qc]} \BibitemShut
  {NoStop}%
\bibitem [{\citenamefont {Cardoso}\ \emph
  {et~al.}(2016{\natexlab{b}})\citenamefont {Cardoso}, \citenamefont {Hopper},
  \citenamefont {Macedo}, \citenamefont {Palenzuela},\ and\ \citenamefont
  {Pani}}]{Cardoso:2016oxy}%
  \BibitemOpen
  \bibfield  {author} {\bibinfo {author} {\bibfnamefont {V.}~\bibnamefont
  {Cardoso}}, \bibinfo {author} {\bibfnamefont {S.}~\bibnamefont {Hopper}},
  \bibinfo {author} {\bibfnamefont {C.~F.~B.}\ \bibnamefont {Macedo}}, \bibinfo
  {author} {\bibfnamefont {C.}~\bibnamefont {Palenzuela}},\ and\ \bibinfo
  {author} {\bibfnamefont {P.}~\bibnamefont {Pani}},\ }\bibfield  {title}
  {\bibinfo {title} {{Gravitational-wave signatures of exotic compact objects
  and of quantum corrections at the horizon scale}},\ }\href
  {https://doi.org/10.1103/PhysRevD.94.084031} {\bibfield  {journal} {\bibinfo
  {journal} {Phys. Rev. D}\ }\textbf {\bibinfo {volume} {94}},\ \bibinfo
  {pages} {084031} (\bibinfo {year} {2016}{\natexlab{b}})},\ \Eprint
  {https://arxiv.org/abs/1608.08637} {arXiv:1608.08637 [gr-qc]} \BibitemShut
  {NoStop}%
\bibitem [{\citenamefont {Carson}\ and\ \citenamefont
  {Yagi}(2020)}]{Carson:2019kkh}%
  \BibitemOpen
  \bibfield  {author} {\bibinfo {author} {\bibfnamefont {Z.}~\bibnamefont
  {Carson}}\ and\ \bibinfo {author} {\bibfnamefont {K.}~\bibnamefont {Yagi}},\
  }\bibfield  {title} {\bibinfo {title} {{Parametrized and
  inspiral-merger-ringdown consistency tests of gravity with multiband
  gravitational wave observations}},\ }\href
  {https://doi.org/10.1103/PhysRevD.101.044047} {\bibfield  {journal} {\bibinfo
   {journal} {Phys. Rev. D}\ }\textbf {\bibinfo {volume} {101}},\ \bibinfo
  {pages} {044047} (\bibinfo {year} {2020})},\ \Eprint
  {https://arxiv.org/abs/1911.05258} {arXiv:1911.05258 [gr-qc]} \BibitemShut
  {NoStop}%
\bibitem [{\citenamefont {Abedi}\ \emph {et~al.}(2020)\citenamefont {Abedi},
  \citenamefont {Afshordi}, \citenamefont {Oshita},\ and\ \citenamefont
  {Wang}}]{Abedi:2020ujo}%
  \BibitemOpen
  \bibfield  {author} {\bibinfo {author} {\bibfnamefont {J.}~\bibnamefont
  {Abedi}}, \bibinfo {author} {\bibfnamefont {N.}~\bibnamefont {Afshordi}},
  \bibinfo {author} {\bibfnamefont {N.}~\bibnamefont {Oshita}},\ and\ \bibinfo
  {author} {\bibfnamefont {Q.}~\bibnamefont {Wang}},\ }\bibfield  {title}
  {\bibinfo {title} {{Quantum Black Holes in the Sky}},\ }\href
  {https://doi.org/10.3390/universe6030043} {\bibfield  {journal} {\bibinfo
  {journal} {Universe}\ }\textbf {\bibinfo {volume} {6}},\ \bibinfo {pages}
  {43} (\bibinfo {year} {2020})},\ \Eprint {https://arxiv.org/abs/2001.09553}
  {arXiv:2001.09553 [gr-qc]} \BibitemShut {NoStop}%
\bibitem [{\citenamefont {Bhagwat}\ and\ \citenamefont
  {Pacilio}(2021)}]{Bhagwat:2021kfa}%
  \BibitemOpen
  \bibfield  {author} {\bibinfo {author} {\bibfnamefont {S.}~\bibnamefont
  {Bhagwat}}\ and\ \bibinfo {author} {\bibfnamefont {C.}~\bibnamefont
  {Pacilio}},\ }\bibfield  {title} {\bibinfo {title} {{Merger-ringdown
  consistency: A new test of strong gravity using deep learning}},\ }\href
  {https://doi.org/10.1103/PhysRevD.104.024030} {\bibfield  {journal} {\bibinfo
   {journal} {Phys. Rev. D}\ }\textbf {\bibinfo {volume} {104}},\ \bibinfo
  {pages} {024030} (\bibinfo {year} {2021})},\ \Eprint
  {https://arxiv.org/abs/2101.07817} {arXiv:2101.07817 [gr-qc]} \BibitemShut
  {NoStop}%
\bibitem [{\citenamefont {Maggiore}\ \emph {et~al.}(2020)\citenamefont
  {Maggiore} \emph {et~al.}}]{Maggiore:2019uih}%
  \BibitemOpen
  \bibfield  {author} {\bibinfo {author} {\bibfnamefont {M.}~\bibnamefont
  {Maggiore}} \emph {et~al.},\ }\bibfield  {title} {\bibinfo {title} {{Science
  Case for the Einstein Telescope}},\ }\href
  {https://doi.org/10.1088/1475-7516/2020/03/050} {\bibfield  {journal}
  {\bibinfo  {journal} {J. Cosmol. Astropart. Phys.}\ }\textbf {\bibinfo
  {volume} {\normalfont03}}\bibfield  {number} {\bibinfo  {number} { (2020)},\
  \bibinfo {pages} {050}},\ }\Eprint {https://arxiv.org/abs/1912.02622}
  {arXiv:1912.02622 [astro-ph.CO]} \BibitemShut {NoStop}%
\bibitem [{\citenamefont {Oshita}\ \emph {et~al.}(2020)\citenamefont {Oshita},
  \citenamefont {Wang},\ and\ \citenamefont {Afshordi}}]{Oshita:2019sat}%
  \BibitemOpen
  \bibfield  {author} {\bibinfo {author} {\bibfnamefont {N.}~\bibnamefont
  {Oshita}}, \bibinfo {author} {\bibfnamefont {Q.}~\bibnamefont {Wang}},\ and\
  \bibinfo {author} {\bibfnamefont {N.}~\bibnamefont {Afshordi}},\ }\bibfield
  {title} {\bibinfo {title} {{On Reflectivity of Quantum Black Hole
  Horizons}},\ }\href {https://doi.org/10.1088/1475-7516/2020/04/016}
  {\bibfield  {journal} {\bibinfo  {journal} {J. Cosmol. Astropart. Phys.}\
  }\textbf {\bibinfo {volume} {\normalfont04}}\bibfield  {number} {\bibinfo
  {number} { (2020)},\ \bibinfo {pages} {016}},\ }\Eprint
  {https://arxiv.org/abs/1905.00464} {arXiv:1905.00464 [hep-th]} \BibitemShut
  {NoStop}%
\bibitem [{\citenamefont {Abedi}\ and\ \citenamefont
  {Afshordi}(2019)}]{Abedi:2018npz}%
  \BibitemOpen
  \bibfield  {author} {\bibinfo {author} {\bibfnamefont {J.}~\bibnamefont
  {Abedi}}\ and\ \bibinfo {author} {\bibfnamefont {N.}~\bibnamefont
  {Afshordi}},\ }\bibfield  {title} {\bibinfo {title} {{Echoes from the Abyss:
  A highly spinning black hole remnant for the binary neutron star merger
  GW170817}},\ }\href {https://doi.org/10.1088/1475-7516/2019/11/010}
  {\bibfield  {journal} {\bibinfo  {journal} {J. Cosmol. Astropart. Phys.}\
  }\textbf {\bibinfo {volume} {\normalfont11}}\bibfield  {number} {\bibinfo
  {number} { (2019)},\ \bibinfo {pages} {010}},\ }\Eprint
  {https://arxiv.org/abs/1803.10454} {arXiv:1803.10454 [gr-qc]} \BibitemShut
  {NoStop}%
\bibitem [{\citenamefont {Dey}\ \emph {et~al.}(2020)\citenamefont {Dey},
  \citenamefont {Chakraborty},\ and\ \citenamefont {Afshordi}}]{Dey:2020lhq}%
  \BibitemOpen
  \bibfield  {author} {\bibinfo {author} {\bibfnamefont {R.}~\bibnamefont
  {Dey}}, \bibinfo {author} {\bibfnamefont {S.}~\bibnamefont {Chakraborty}},\
  and\ \bibinfo {author} {\bibfnamefont {N.}~\bibnamefont {Afshordi}},\
  }\bibfield  {title} {\bibinfo {title} {{Echoes from braneworld black
  holes}},\ }\href {https://doi.org/10.1103/PhysRevD.101.104014} {\bibfield
  {journal} {\bibinfo  {journal} {Phys. Rev. D}\ }\textbf {\bibinfo {volume}
  {101}},\ \bibinfo {pages} {104014} (\bibinfo {year} {2020})},\ \Eprint
  {https://arxiv.org/abs/2001.01301} {arXiv:2001.01301 [gr-qc]} \BibitemShut
  {NoStop}%
\bibitem [{\citenamefont {Bueno}\ \emph {et~al.}(2018)\citenamefont {Bueno},
  \citenamefont {Cano}, \citenamefont {Goelen}, \citenamefont {Hertog},\ and\
  \citenamefont {Vercnocke}}]{Bueno:2017hyj}%
  \BibitemOpen
  \bibfield  {author} {\bibinfo {author} {\bibfnamefont {P.}~\bibnamefont
  {Bueno}}, \bibinfo {author} {\bibfnamefont {P.~A.}\ \bibnamefont {Cano}},
  \bibinfo {author} {\bibfnamefont {F.}~\bibnamefont {Goelen}}, \bibinfo
  {author} {\bibfnamefont {T.}~\bibnamefont {Hertog}},\ and\ \bibinfo {author}
  {\bibfnamefont {B.}~\bibnamefont {Vercnocke}},\ }\bibfield  {title} {\bibinfo
  {title} {{Echoes of Kerr-like wormholes}},\ }\href
  {https://doi.org/10.1103/PhysRevD.97.024040} {\bibfield  {journal} {\bibinfo
  {journal} {Phys. Rev. D}\ }\textbf {\bibinfo {volume} {97}},\ \bibinfo
  {pages} {024040} (\bibinfo {year} {2018})},\ \Eprint
  {https://arxiv.org/abs/1711.00391} {arXiv:1711.00391 [gr-qc]} \BibitemShut
  {NoStop}%
\bibitem [{\citenamefont {Mark}\ \emph {et~al.}(2017)\citenamefont {Mark},
  \citenamefont {Zimmerman}, \citenamefont {Du},\ and\ \citenamefont
  {Chen}}]{Mark:2017dnq}%
  \BibitemOpen
  \bibfield  {author} {\bibinfo {author} {\bibfnamefont {Z.}~\bibnamefont
  {Mark}}, \bibinfo {author} {\bibfnamefont {A.}~\bibnamefont {Zimmerman}},
  \bibinfo {author} {\bibfnamefont {S.~M.}\ \bibnamefont {Du}},\ and\ \bibinfo
  {author} {\bibfnamefont {Y.}~\bibnamefont {Chen}},\ }\bibfield  {title}
  {\bibinfo {title} {{A recipe for echoes from exotic compact objects}},\
  }\href {https://doi.org/10.1103/PhysRevD.96.084002} {\bibfield  {journal}
  {\bibinfo  {journal} {Phys. Rev. D}\ }\textbf {\bibinfo {volume} {96}},\
  \bibinfo {pages} {084002} (\bibinfo {year} {2017})},\ \Eprint
  {https://arxiv.org/abs/1706.06155} {arXiv:1706.06155 [gr-qc]} \BibitemShut
  {NoStop}%
\bibitem [{\citenamefont {Konoplya}\ and\ \citenamefont
  {Zhidenko}(2016)}]{Konoplya:2016pmh}%
  \BibitemOpen
  \bibfield  {author} {\bibinfo {author} {\bibfnamefont {R.}~\bibnamefont
  {Konoplya}}\ and\ \bibinfo {author} {\bibfnamefont {A.}~\bibnamefont
  {Zhidenko}},\ }\bibfield  {title} {\bibinfo {title} {{Detection of
  gravitational waves from black holes: Is there a window for alternative
  theories?}},\ }\href {https://doi.org/10.1016/j.physletb.2016.03.044}
  {\bibfield  {journal} {\bibinfo  {journal} {Phys. Lett. B}\ }\textbf
  {\bibinfo {volume} {756}},\ \bibinfo {pages} {350} (\bibinfo {year}
  {2016})},\ \Eprint {https://arxiv.org/abs/1602.04738} {arXiv:1602.04738
  [gr-qc]} \BibitemShut {NoStop}%
\bibitem [{\citenamefont {Press}\ and\ \citenamefont
  {Teukolsky}(1972)}]{Press:1972zz}%
  \BibitemOpen
  \bibfield  {author} {\bibinfo {author} {\bibfnamefont {W.~H.}\ \bibnamefont
  {Press}}\ and\ \bibinfo {author} {\bibfnamefont {S.~A.}\ \bibnamefont
  {Teukolsky}},\ }\bibfield  {title} {\bibinfo {title} {{Floating Orbits,
  Superradiant Scattering and the Black-hole Bomb}},\ }\href
  {https://doi.org/10.1038/238211a0} {\bibfield  {journal} {\bibinfo  {journal}
  {Nature}\ }\textbf {\bibinfo {volume} {238}},\ \bibinfo {pages} {211}
  (\bibinfo {year} {1972})}\BibitemShut {NoStop}%
\bibitem [{\citenamefont {Cardoso}\ \emph {et~al.}(2004)\citenamefont
  {Cardoso}, \citenamefont {Dias}, \citenamefont {Lemos},\ and\ \citenamefont
  {Yoshida}}]{Cardoso:2004nk}%
  \BibitemOpen
  \bibfield  {author} {\bibinfo {author} {\bibfnamefont {V.}~\bibnamefont
  {Cardoso}}, \bibinfo {author} {\bibfnamefont {O.~J.~C.}\ \bibnamefont
  {Dias}}, \bibinfo {author} {\bibfnamefont {J.~P.~S.}\ \bibnamefont {Lemos}},\
  and\ \bibinfo {author} {\bibfnamefont {S.}~\bibnamefont {Yoshida}},\
  }\bibfield  {title} {\bibinfo {title} {{The black hole bomb and superradiant
  instabilities}},\ }\href {https://doi.org/10.1103/PhysRevD.70.044039}
  {\bibfield  {journal} {\bibinfo  {journal} {Phys. Rev. D}\ }\textbf {\bibinfo
  {volume} {70}},\ \bibinfo {pages} {044039} (\bibinfo {year} {2004})},\
  \bibinfo {note} {[Erratum: Phys. Rev. D {\bf70}, 049903(E) (2004)]},\ \Eprint
  {https://arxiv.org/abs/hep-th/0404096} {arXiv:hep-th/0404096} \BibitemShut
  {NoStop}%
\bibitem [{\citenamefont {Hod}(2017)}]{Hod:2017cga}%
  \BibitemOpen
  \bibfield  {author} {\bibinfo {author} {\bibfnamefont {S.}~\bibnamefont
  {Hod}},\ }\bibfield  {title} {\bibinfo {title} {{Onset of superradiant
  instabilities in rotating spacetimes of exotic compact objects}},\ }\href
  {https://doi.org/10.1007/JHEP06(2017)132} {\bibfield  {journal} {\bibinfo
  {journal} {J. High Energy Phys.}\ }\textbf {\bibinfo {volume}
  {\normalfont06}}\bibfield  {number} {\bibinfo  {number} { (2017)},\ \bibinfo
  {pages} {132}},\ }\Eprint {https://arxiv.org/abs/1704.05856}
  {arXiv:1704.05856 [hep-th]} \BibitemShut {NoStop}%
\bibitem [{\citenamefont {Addazi}\ \emph {et~al.}(2020)\citenamefont {Addazi},
  \citenamefont {Marcian\`o},\ and\ \citenamefont {Yunes}}]{Addazi:2019bjz}%
  \BibitemOpen
  \bibfield  {author} {\bibinfo {author} {\bibfnamefont {A.}~\bibnamefont
  {Addazi}}, \bibinfo {author} {\bibfnamefont {A.}~\bibnamefont {Marcian\`o}},\
  and\ \bibinfo {author} {\bibfnamefont {N.}~\bibnamefont {Yunes}},\ }\bibfield
   {title} {\bibinfo {title} {{Gravitational Instability of Exotic Compact
  Objects}},\ }\href {https://doi.org/10.1140/epjc/s10052-019-7575-9}
  {\bibfield  {journal} {\bibinfo  {journal} {Eur. Phys. J. C}\ }\textbf
  {\bibinfo {volume} {80}},\ \bibinfo {pages} {36} (\bibinfo {year} {2020})},\
  \Eprint {https://arxiv.org/abs/1905.08734} {arXiv:1905.08734 [gr-qc]}
  \BibitemShut {NoStop}%
\bibitem [{\citenamefont {Brito}\ \emph {et~al.}(2020)\citenamefont {Brito},
  \citenamefont {Cardoso},\ and\ \citenamefont
  {Pani}}]{brito_superradiance_2020}%
  \BibitemOpen
  \bibfield  {author} {\bibinfo {author} {\bibfnamefont {R.}~\bibnamefont
  {Brito}}, \bibinfo {author} {\bibfnamefont {V.}~\bibnamefont {Cardoso}},\
  and\ \bibinfo {author} {\bibfnamefont {P.}~\bibnamefont {Pani}},\ }\href
  {https://doi.org/10.1007/978-3-030-46622-0} {\emph {\bibinfo {title}
  {{Superradiance:\ New Frontiers in Black Hole Physics}}}},\ \bibinfo
  {edition} {2nd}\ ed.,\ \bibinfo {series} {Lect. Notes Phys.}, Vol.\ \bibinfo
  {volume} {971}\ (\bibinfo  {publisher} {Springer},\ \bibinfo {year} {2020})\
  \Eprint {https://arxiv.org/abs/1501.06570} {arXiv:1501.06570 [gr-qc]}
  \BibitemShut {NoStop}%
\bibitem [{\citenamefont {Franzin}\ \emph
  {et~al.}(2021{\natexlab{a}})\citenamefont {Franzin}, \citenamefont
  {Liberati},\ and\ \citenamefont {Oi}}]{Franzin:2021kvj}%
  \BibitemOpen
  \bibfield  {author} {\bibinfo {author} {\bibfnamefont {E.}~\bibnamefont
  {Franzin}}, \bibinfo {author} {\bibfnamefont {S.}~\bibnamefont {Liberati}},\
  and\ \bibinfo {author} {\bibfnamefont {M.}~\bibnamefont {Oi}},\ }\bibfield
  {title} {\bibinfo {title} {{Superradiance in Kerr-like black holes}},\ }\href
  {https://doi.org/10.1103/PhysRevD.103.104034} {\bibfield  {journal} {\bibinfo
   {journal} {Phys. Rev. D}\ }\textbf {\bibinfo {volume} {103}},\ \bibinfo
  {pages} {104034} (\bibinfo {year} {2021}{\natexlab{a}})},\ \Eprint
  {https://arxiv.org/abs/2102.03152} {arXiv:2102.03152 [gr-qc]} \BibitemShut
  {NoStop}%
\bibitem [{\citenamefont {Cardoso}\ \emph
  {et~al.}(2008{\natexlab{a}})\citenamefont {Cardoso}, \citenamefont {Pani},
  \citenamefont {Cadoni},\ and\ \citenamefont {Cavagli\`a}}]{Cardoso:2007az}%
  \BibitemOpen
  \bibfield  {author} {\bibinfo {author} {\bibfnamefont {V.}~\bibnamefont
  {Cardoso}}, \bibinfo {author} {\bibfnamefont {P.}~\bibnamefont {Pani}},
  \bibinfo {author} {\bibfnamefont {M.}~\bibnamefont {Cadoni}},\ and\ \bibinfo
  {author} {\bibfnamefont {M.}~\bibnamefont {Cavagli\`a}},\ }\bibfield  {title}
  {\bibinfo {title} {{Ergoregion instability of ultracompact astrophysical
  objects}},\ }\href {https://doi.org/10.1103/PhysRevD.77.124044} {\bibfield
  {journal} {\bibinfo  {journal} {Phys. Rev. D}\ }\textbf {\bibinfo {volume}
  {77}},\ \bibinfo {pages} {124044} (\bibinfo {year} {2008}{\natexlab{a}})},\
  \Eprint {https://arxiv.org/abs/0709.0532} {arXiv:0709.0532 [gr-qc]}
  \BibitemShut {NoStop}%
\bibitem [{\citenamefont {Cardoso}\ \emph
  {et~al.}(2008{\natexlab{b}})\citenamefont {Cardoso}, \citenamefont {Pani},
  \citenamefont {Cadoni},\ and\ \citenamefont {Cavagli\`a}}]{Cardoso:2008kj}%
  \BibitemOpen
  \bibfield  {author} {\bibinfo {author} {\bibfnamefont {V.}~\bibnamefont
  {Cardoso}}, \bibinfo {author} {\bibfnamefont {P.}~\bibnamefont {Pani}},
  \bibinfo {author} {\bibfnamefont {M.}~\bibnamefont {Cadoni}},\ and\ \bibinfo
  {author} {\bibfnamefont {M.}~\bibnamefont {Cavagli\`a}},\ }\bibfield  {title}
  {\bibinfo {title} {{Instability of hyper-compact Kerr-like objects}},\ }\href
  {https://doi.org/10.1088/0264-9381/25/19/195010} {\bibfield  {journal}
  {\bibinfo  {journal} {Class. Quantum Grav.}\ }\textbf {\bibinfo {volume}
  {25}},\ \bibinfo {pages} {195010} (\bibinfo {year} {2008}{\natexlab{b}})},\
  \Eprint {https://arxiv.org/abs/0808.1615} {arXiv:0808.1615 [gr-qc]}
  \BibitemShut {NoStop}%
\bibitem [{\citenamefont {Pani}\ \emph {et~al.}(2010)\citenamefont {Pani},
  \citenamefont {Barausse}, \citenamefont {Berti},\ and\ \citenamefont
  {Cardoso}}]{Pani:2010jz}%
  \BibitemOpen
  \bibfield  {author} {\bibinfo {author} {\bibfnamefont {P.}~\bibnamefont
  {Pani}}, \bibinfo {author} {\bibfnamefont {E.}~\bibnamefont {Barausse}},
  \bibinfo {author} {\bibfnamefont {E.}~\bibnamefont {Berti}},\ and\ \bibinfo
  {author} {\bibfnamefont {V.}~\bibnamefont {Cardoso}},\ }\bibfield  {title}
  {\bibinfo {title} {{Gravitational instabilities of superspinars}},\ }\href
  {https://doi.org/10.1103/PhysRevD.82.044009} {\bibfield  {journal} {\bibinfo
  {journal} {Phys. Rev. D}\ }\textbf {\bibinfo {volume} {82}},\ \bibinfo
  {pages} {044009} (\bibinfo {year} {2010})},\ \Eprint
  {https://arxiv.org/abs/1006.1863} {arXiv:1006.1863 [gr-qc]} \BibitemShut
  {NoStop}%
\bibitem [{\citenamefont {Maggio}\ \emph {et~al.}(2017)\citenamefont {Maggio},
  \citenamefont {Pani},\ and\ \citenamefont {Ferrari}}]{Maggio:2017ivp}%
  \BibitemOpen
  \bibfield  {author} {\bibinfo {author} {\bibfnamefont {E.}~\bibnamefont
  {Maggio}}, \bibinfo {author} {\bibfnamefont {P.}~\bibnamefont {Pani}},\ and\
  \bibinfo {author} {\bibfnamefont {V.}~\bibnamefont {Ferrari}},\ }\bibfield
  {title} {\bibinfo {title} {{Exotic Compact Objects and How to Quench their
  Ergoregion Instability}},\ }\href
  {https://doi.org/10.1103/PhysRevD.96.104047} {\bibfield  {journal} {\bibinfo
  {journal} {Phys. Rev. D}\ }\textbf {\bibinfo {volume} {96}},\ \bibinfo
  {pages} {104047} (\bibinfo {year} {2017})},\ \Eprint
  {https://arxiv.org/abs/1703.03696} {arXiv:1703.03696 [gr-qc]} \BibitemShut
  {NoStop}%
\bibitem [{\citenamefont {Maggio}\ \emph {et~al.}(2019)\citenamefont {Maggio},
  \citenamefont {Cardoso}, \citenamefont {Dolan},\ and\ \citenamefont
  {Pani}}]{Maggio:2018ivz}%
  \BibitemOpen
  \bibfield  {author} {\bibinfo {author} {\bibfnamefont {E.}~\bibnamefont
  {Maggio}}, \bibinfo {author} {\bibfnamefont {V.}~\bibnamefont {Cardoso}},
  \bibinfo {author} {\bibfnamefont {S.~R.}\ \bibnamefont {Dolan}},\ and\
  \bibinfo {author} {\bibfnamefont {P.}~\bibnamefont {Pani}},\ }\bibfield
  {title} {\bibinfo {title} {{Ergoregion instability of exotic compact objects:
  electromagnetic and gravitational perturbations and the role of
  absorption}},\ }\href {https://doi.org/10.1103/PhysRevD.99.064007} {\bibfield
   {journal} {\bibinfo  {journal} {Phys. Rev. D}\ }\textbf {\bibinfo {volume}
  {99}},\ \bibinfo {pages} {064007} (\bibinfo {year} {2019})},\ \Eprint
  {https://arxiv.org/abs/1807.08840} {arXiv:1807.08840 [gr-qc]} \BibitemShut
  {NoStop}%
\bibitem [{\citenamefont {Dey}\ \emph {et~al.}(2021)\citenamefont {Dey},
  \citenamefont {Biswas},\ and\ \citenamefont {Chakraborty}}]{Dey:2020pth}%
  \BibitemOpen
  \bibfield  {author} {\bibinfo {author} {\bibfnamefont {R.}~\bibnamefont
  {Dey}}, \bibinfo {author} {\bibfnamefont {S.}~\bibnamefont {Biswas}},\ and\
  \bibinfo {author} {\bibfnamefont {S.}~\bibnamefont {Chakraborty}},\
  }\bibfield  {title} {\bibinfo {title} {{Ergoregion instability and echoes for
  braneworld black holes: Scalar, electromagnetic, and gravitational
  perturbations}},\ }\href {https://doi.org/10.1103/PhysRevD.103.084019}
  {\bibfield  {journal} {\bibinfo  {journal} {Phys. Rev. D}\ }\textbf {\bibinfo
  {volume} {103}},\ \bibinfo {pages} {084019} (\bibinfo {year} {2021})},\
  \Eprint {https://arxiv.org/abs/2010.07966} {arXiv:2010.07966 [gr-qc]}
  \BibitemShut {NoStop}%
\bibitem [{\citenamefont {Carballo-Rubio}\ \emph {et~al.}(2018)\citenamefont
  {Carballo-Rubio}, \citenamefont {Di~Filippo}, \citenamefont {Liberati},
  \citenamefont {Pacilio},\ and\ \citenamefont
  {Visser}}]{Carballo-Rubio:2018pmi}%
  \BibitemOpen
  \bibfield  {author} {\bibinfo {author} {\bibfnamefont {R.}~\bibnamefont
  {Carballo-Rubio}}, \bibinfo {author} {\bibfnamefont {F.}~\bibnamefont
  {Di~Filippo}}, \bibinfo {author} {\bibfnamefont {S.}~\bibnamefont
  {Liberati}}, \bibinfo {author} {\bibfnamefont {C.}~\bibnamefont {Pacilio}},\
  and\ \bibinfo {author} {\bibfnamefont {M.}~\bibnamefont {Visser}},\
  }\bibfield  {title} {\bibinfo {title} {{On the viability of regular black
  holes}},\ }\href {https://doi.org/10.1007/JHEP07(2018)023} {\bibfield
  {journal} {\bibinfo  {journal} {J. High Energy Phys.}\ }\textbf {\bibinfo
  {volume} {\normalfont07}}\bibfield  {number} {\bibinfo  {number} { (2018)},\
  \bibinfo {pages} {023}},\ }\Eprint {https://arxiv.org/abs/1805.02675}
  {arXiv:1805.02675 [gr-qc]} \BibitemShut {NoStop}%
\bibitem [{\citenamefont {Carballo-Rubio}\ \emph {et~al.}(2021)\citenamefont
  {Carballo-Rubio}, \citenamefont {Di~Filippo}, \citenamefont {Liberati},
  \citenamefont {Pacilio},\ and\ \citenamefont
  {Visser}}]{Carballo-Rubio:2021bpr}%
  \BibitemOpen
  \bibfield  {author} {\bibinfo {author} {\bibfnamefont {R.}~\bibnamefont
  {Carballo-Rubio}}, \bibinfo {author} {\bibfnamefont {F.}~\bibnamefont
  {Di~Filippo}}, \bibinfo {author} {\bibfnamefont {S.}~\bibnamefont
  {Liberati}}, \bibinfo {author} {\bibfnamefont {C.}~\bibnamefont {Pacilio}},\
  and\ \bibinfo {author} {\bibfnamefont {M.}~\bibnamefont {Visser}},\
  }\bibfield  {title} {\bibinfo {title} {{Inner horizon instability and the
  unstable cores of regular black holes}},\ }\href
  {https://doi.org/10.1007/JHEP05(2021)132} {\bibfield  {journal} {\bibinfo
  {journal} {J. High Energy Phys.}\ }\textbf {\bibinfo {volume}
  {\normalfont05}}\bibfield  {number} {\bibinfo  {number} { (2021)},\ \bibinfo
  {pages} {132}},\ }\Eprint {https://arxiv.org/abs/2101.05006}
  {arXiv:2101.05006 [gr-qc]} \BibitemShut {NoStop}%
\bibitem [{\citenamefont {Di~Filippo}(2020)}]{DiFilippo:2020ooa}%
  \BibitemOpen
  \bibfield  {author} {\bibinfo {author} {\bibfnamefont {F.}~\bibnamefont
  {Di~Filippo}},\ }\emph {\bibinfo {title} {{Beyond General Relativity:
  Modified Theories and Non-Singular Black Holes}}},\ \href
  {https://inspirehep.net/files/46fce2d67a7e827dcfa81acc6890fc51} {Ph.D.
  thesis},\ \bibinfo  {school} {SISSA, Trieste} (\bibinfo {year}
  {2020})\BibitemShut {NoStop}%
\bibitem [{\citenamefont {Cardoso}\ \emph {et~al.}(2018)\citenamefont
  {Cardoso}, \citenamefont {Costa}, \citenamefont {Destounis}, \citenamefont
  {Hintz},\ and\ \citenamefont {Jansen}}]{Cardoso:2017soq}%
  \BibitemOpen
  \bibfield  {author} {\bibinfo {author} {\bibfnamefont {V.}~\bibnamefont
  {Cardoso}}, \bibinfo {author} {\bibfnamefont {J.~L.}\ \bibnamefont {Costa}},
  \bibinfo {author} {\bibfnamefont {K.}~\bibnamefont {Destounis}}, \bibinfo
  {author} {\bibfnamefont {P.}~\bibnamefont {Hintz}},\ and\ \bibinfo {author}
  {\bibfnamefont {A.}~\bibnamefont {Jansen}},\ }\bibfield  {title} {\bibinfo
  {title} {{Quasinormal modes and Strong Cosmic Censorship}},\ }\href
  {https://doi.org/10.1103/PhysRevLett.120.031103} {\bibfield  {journal}
  {\bibinfo  {journal} {Phys. Rev. Lett.}\ }\textbf {\bibinfo {volume} {120}},\
  \bibinfo {pages} {031103} (\bibinfo {year} {2018})},\ \Eprint
  {https://arxiv.org/abs/1711.10502} {arXiv:1711.10502 [gr-qc]} \BibitemShut
  {NoStop}%
\bibitem [{\citenamefont {Dias}\ \emph {et~al.}(2018)\citenamefont {Dias},
  \citenamefont {Eperon}, \citenamefont {Reall},\ and\ \citenamefont
  {Santos}}]{Dias:2018ynt}%
  \BibitemOpen
  \bibfield  {author} {\bibinfo {author} {\bibfnamefont {O.~J.~C.}\
  \bibnamefont {Dias}}, \bibinfo {author} {\bibfnamefont {F.~C.}\ \bibnamefont
  {Eperon}}, \bibinfo {author} {\bibfnamefont {H.~S.}\ \bibnamefont {Reall}},\
  and\ \bibinfo {author} {\bibfnamefont {J.~E.}\ \bibnamefont {Santos}},\
  }\bibfield  {title} {\bibinfo {title} {{Strong cosmic censorship in de Sitter
  space}},\ }\href {https://doi.org/10.1103/PhysRevD.97.104060} {\bibfield
  {journal} {\bibinfo  {journal} {Phys. Rev. D}\ }\textbf {\bibinfo {volume}
  {97}},\ \bibinfo {pages} {104060} (\bibinfo {year} {2018})},\ \Eprint
  {https://arxiv.org/abs/1801.09694} {arXiv:1801.09694 [gr-qc]} \BibitemShut
  {NoStop}%
\bibitem [{\citenamefont {Mishra}\ and\ \citenamefont
  {Chakraborty}(2020)}]{Mishra:2020jlw}%
  \BibitemOpen
  \bibfield  {author} {\bibinfo {author} {\bibfnamefont {A.~K.}\ \bibnamefont
  {Mishra}}\ and\ \bibinfo {author} {\bibfnamefont {S.}~\bibnamefont
  {Chakraborty}},\ }\bibfield  {title} {\bibinfo {title} {{Strong cosmic
  censorship conjecture in higher curvature gravity}},\ }\href
  {https://doi.org/10.1103/PhysRevD.101.064041} {\bibfield  {journal} {\bibinfo
   {journal} {Phys. Rev. D}\ }\textbf {\bibinfo {volume} {101}},\ \bibinfo
  {pages} {064041} (\bibinfo {year} {2020})},\ \Eprint
  {https://arxiv.org/abs/1911.09855} {arXiv:1911.09855 [gr-qc]} \BibitemShut
  {NoStop}%
\bibitem [{\citenamefont {Rahman}\ \emph {et~al.}(2020)\citenamefont {Rahman},
  \citenamefont {Mitra},\ and\ \citenamefont {Chakraborty}}]{Rahman:2020guv}%
  \BibitemOpen
  \bibfield  {author} {\bibinfo {author} {\bibfnamefont {M.}~\bibnamefont
  {Rahman}}, \bibinfo {author} {\bibfnamefont {S.}~\bibnamefont {Mitra}},\ and\
  \bibinfo {author} {\bibfnamefont {S.}~\bibnamefont {Chakraborty}},\
  }\bibfield  {title} {\bibinfo {title} {{Strong cosmic censorship conjecture
  with NUT charge and conformal coupling}},\ }\href
  {https://doi.org/10.1088/1361-6382/aba17d} {\bibfield  {journal} {\bibinfo
  {journal} {Class. Quantum Grav.}\ }\textbf {\bibinfo {volume} {37}},\
  \bibinfo {pages} {195004} (\bibinfo {year} {2020})},\ \Eprint
  {https://arxiv.org/abs/2001.00599} {arXiv:2001.00599 [gr-qc]} \BibitemShut
  {NoStop}%
\bibitem [{\citenamefont {Rahman}\ \emph {et~al.}(2019)\citenamefont {Rahman},
  \citenamefont {Chakraborty}, \citenamefont {SenGupta},\ and\ \citenamefont
  {Sen}}]{Rahman:2018oso}%
  \BibitemOpen
  \bibfield  {author} {\bibinfo {author} {\bibfnamefont {M.}~\bibnamefont
  {Rahman}}, \bibinfo {author} {\bibfnamefont {S.}~\bibnamefont {Chakraborty}},
  \bibinfo {author} {\bibfnamefont {S.}~\bibnamefont {SenGupta}},\ and\
  \bibinfo {author} {\bibfnamefont {A.~A.}\ \bibnamefont {Sen}},\ }\bibfield
  {title} {\bibinfo {title} {{Fate of Strong Cosmic Censorship Conjecture in
  Presence of Higher Spacetime Dimensions}},\ }\href
  {https://doi.org/10.1007/JHEP03(2019)178} {\bibfield  {journal} {\bibinfo
  {journal} {J. High Energy Phys.}\ }\textbf {\bibinfo {volume} {03}}\bibfield
  {number} {\bibinfo  {number} { (2019)},\ \bibinfo {pages} {178}},\ }\Eprint
  {https://arxiv.org/abs/1811.08538} {arXiv:1811.08538 [gr-qc]} \BibitemShut
  {NoStop}%
\bibitem [{\citenamefont {Mazza}\ \emph {et~al.}(2021)\citenamefont {Mazza},
  \citenamefont {Franzin},\ and\ \citenamefont {Liberati}}]{Mazza:2021rgq}%
  \BibitemOpen
  \bibfield  {author} {\bibinfo {author} {\bibfnamefont {J.}~\bibnamefont
  {Mazza}}, \bibinfo {author} {\bibfnamefont {E.}~\bibnamefont {Franzin}},\
  and\ \bibinfo {author} {\bibfnamefont {S.}~\bibnamefont {Liberati}},\
  }\bibfield  {title} {\bibinfo {title} {{A novel family of rotating black hole
  mimickers}},\ }\href {https://doi.org/10.1088/1475-7516/2021/04/082}
  {\bibfield  {journal} {\bibinfo  {journal} {J. Cosmol. Astropart. Phys.}\
  }\textbf {\bibinfo {volume} {\normalfont04}}\bibfield  {number} {\bibinfo
  {number} { (2021)},\ \bibinfo {pages} {082}},\ }\Eprint
  {https://arxiv.org/abs/2102.01105} {arXiv:2102.01105 [gr-qc]} \BibitemShut
  {NoStop}%
\bibitem [{\citenamefont {Bambhaniya}\ \emph {et~al.}(2022)\citenamefont
  {Bambhaniya}, \citenamefont {{Saurabh K}}, \citenamefont {Jusufi},\ and\
  \citenamefont {Joshi}}]{bambhaniya_thin_2021}%
  \BibitemOpen
  \bibfield  {author} {\bibinfo {author} {\bibfnamefont {P.}~\bibnamefont
  {Bambhaniya}}, \bibinfo {author} {\bibnamefont {{Saurabh K}}}, \bibinfo
  {author} {\bibfnamefont {K.}~\bibnamefont {Jusufi}},\ and\ \bibinfo {author}
  {\bibfnamefont {P.~S.}\ \bibnamefont {Joshi}},\ }\bibfield  {title} {\bibinfo
  {title} {{Thin accretion disk in the Simpson--Visser black-bounce and
  wormhole spacetimes}},\ }\href {https://doi.org/10.1103/PhysRevD.105.023021}
  {\bibfield  {journal} {\bibinfo  {journal} {Phys. Rev. D}\ }\textbf {\bibinfo
  {volume} {105}},\ \bibinfo {pages} {023021} (\bibinfo {year} {2022})},\
  \Eprint {https://arxiv.org/abs/2109.15054} {arXiv:2109.15054 [gr-qc]}
  \BibitemShut {NoStop}%
\bibitem [{\citenamefont {Guerrero}\ \emph {et~al.}(2021)\citenamefont
  {Guerrero}, \citenamefont {Olmo}, \citenamefont {Rubiera-Garcia},\ and\
  \citenamefont {G\'omez}}]{guerrero_shadows_2021}%
  \BibitemOpen
  \bibfield  {author} {\bibinfo {author} {\bibfnamefont {M.}~\bibnamefont
  {Guerrero}}, \bibinfo {author} {\bibfnamefont {G.~J.}\ \bibnamefont {Olmo}},
  \bibinfo {author} {\bibfnamefont {D.}~\bibnamefont {Rubiera-Garcia}},\ and\
  \bibinfo {author} {\bibfnamefont {D.~S.-C.}\ \bibnamefont {G\'omez}},\
  }\bibfield  {title} {\bibinfo {title} {{Shadows and optical appearance of
  black bounces illuminated by a thin accretion disk}},\ }\href
  {https://doi.org/10.1088/1475-7516/2021/08/036} {\bibfield  {journal}
  {\bibinfo  {journal} {J. Cosmol. Astropart. Phys.}\ }\textbf {\bibinfo
  {volume} {\normalfont08}}\bibfield  {number} {\bibinfo  {number} { (2021)},\
  \bibinfo {pages} {036}},\ }\Eprint {https://arxiv.org/abs/2105.15073}
  {arXiv:2105.15073 [gr-qc]} \BibitemShut {NoStop}%
\bibitem [{\citenamefont {Islam}\ \emph {et~al.}(2021)\citenamefont {Islam},
  \citenamefont {Kumar},\ and\ \citenamefont {Ghosh}}]{Islam:2021ful}%
  \BibitemOpen
  \bibfield  {author} {\bibinfo {author} {\bibfnamefont {S.~U.}\ \bibnamefont
  {Islam}}, \bibinfo {author} {\bibfnamefont {J.}~\bibnamefont {Kumar}},\ and\
  \bibinfo {author} {\bibfnamefont {S.~G.}\ \bibnamefont {Ghosh}},\ }\bibfield
  {title} {\bibinfo {title} {{Strong gravitational lensing by rotating
  Simpson--Visser black holes}},\ }\href
  {https://doi.org/10.1088/1475-7516/2021/10/013} {\bibfield  {journal}
  {\bibinfo  {journal} {J. Cosmol. Astropart. Phys.}\ }\textbf {\bibinfo
  {volume} {\normalfont10}}\bibfield  {number} {\bibinfo  {number} { (2021)},\
  \bibinfo {pages} {013}},\ }\Eprint {https://arxiv.org/abs/2104.00696}
  {arXiv:2104.00696 [gr-qc]} \BibitemShut {NoStop}%
\bibitem [{\citenamefont {Jafarzade}\ \emph {et~al.}(2021)\citenamefont
  {Jafarzade}, \citenamefont {Zangeneh},\ and\ \citenamefont
  {Lobo}}]{jafarzade_observational_2021}%
  \BibitemOpen
  \bibfield  {author} {\bibinfo {author} {\bibfnamefont {K.}~\bibnamefont
  {Jafarzade}}, \bibinfo {author} {\bibfnamefont {M.~K.}\ \bibnamefont
  {Zangeneh}},\ and\ \bibinfo {author} {\bibfnamefont {F.~S.~N.}\ \bibnamefont
  {Lobo}},\ }\href@noop {} {\bibinfo {title} {{Observational Optical
  Constraints of the Simpson--Visser Black-Bounce Geometry}}} (\bibinfo {year}
  {2021}),\ \Eprint {https://arxiv.org/abs/2106.13893} {arXiv:2106.13893}
  \BibitemShut {NoStop}%
\bibitem [{\citenamefont {{Lima Junior}}\ \emph {et~al.}(2021)\citenamefont
  {{Lima Junior}}, \citenamefont {{Crispino}}, \citenamefont {{Cunha}},\ and\
  \citenamefont {{Herdeiro}}}]{lima_junior_mistaken_2021}%
  \BibitemOpen
  \bibfield  {author} {\bibinfo {author} {\bibfnamefont {H.~C.~D.}\
  \bibnamefont {{Lima Junior}}}, \bibinfo {author} {\bibfnamefont {L.~C.~B.}\
  \bibnamefont {{Crispino}}}, \bibinfo {author} {\bibfnamefont {P.~V.~P.}\
  \bibnamefont {{Cunha}}},\ and\ \bibinfo {author} {\bibfnamefont {C.~A.~R.}\
  \bibnamefont {{Herdeiro}}},\ }\bibfield  {title} {\bibinfo {title} {{Can
  different black holes cast the same shadow?}},\ }\href
  {https://doi.org/10.1103/PhysRevD.103.084040} {\bibfield  {journal} {\bibinfo
   {journal} {Phys. Rev. D}\ }\textbf {\bibinfo {volume} {103}},\ \bibinfo
  {eid} {084040} (\bibinfo {year} {2021})},\ \Eprint
  {https://arxiv.org/abs/2102.07034} {arXiv:2102.07034 [gr-qc]} \BibitemShut
  {NoStop}%
\bibitem [{\citenamefont {Shaikh}\ \emph {et~al.}(2021)\citenamefont {Shaikh},
  \citenamefont {Pal}, \citenamefont {Pal},\ and\ \citenamefont
  {Sarkar}}]{shaikh_constraining_2021}%
  \BibitemOpen
  \bibfield  {author} {\bibinfo {author} {\bibfnamefont {R.}~\bibnamefont
  {Shaikh}}, \bibinfo {author} {\bibfnamefont {K.}~\bibnamefont {Pal}},
  \bibinfo {author} {\bibfnamefont {K.}~\bibnamefont {Pal}},\ and\ \bibinfo
  {author} {\bibfnamefont {T.}~\bibnamefont {Sarkar}},\ }\bibfield  {title}
  {\bibinfo {title} {{Constraining alternatives to the Kerr black hole}},\
  }\href {https://doi.org/10.1093/mnras/stab1779} {\bibfield  {journal}
  {\bibinfo  {journal} {Mon. Not. Roy. Astron. Soc.}\ }\textbf {\bibinfo
  {volume} {506}},\ \bibinfo {pages} {1229} (\bibinfo {year} {2021})},\ \Eprint
  {https://arxiv.org/abs/2102.04299} {arXiv:2102.04299 [gr-qc]} \BibitemShut
  {NoStop}%
\bibitem [{\citenamefont {Simpson}\ \emph {et~al.}(2019)\citenamefont
  {Simpson}, \citenamefont {Mart\'in-Moruno},\ and\ \citenamefont
  {Visser}}]{simpson_vaidya_2019}%
  \BibitemOpen
  \bibfield  {author} {\bibinfo {author} {\bibfnamefont {A.}~\bibnamefont
  {Simpson}}, \bibinfo {author} {\bibfnamefont {P.}~\bibnamefont
  {Mart\'in-Moruno}},\ and\ \bibinfo {author} {\bibfnamefont {M.}~\bibnamefont
  {Visser}},\ }\bibfield  {title} {\bibinfo {title} {{Vaidya Spacetimes,
  Black-Bounces, and Traversable Wormholes}},\ }\href
  {https://doi.org/10.1088/1361-6382/ab28a5} {\bibfield  {journal} {\bibinfo
  {journal} {Class. Quantum Grav.}\ }\textbf {\bibinfo {volume} {36}},\
  \bibinfo {pages} {145007} (\bibinfo {year} {2019})},\ \Eprint
  {https://arxiv.org/abs/1902.04232} {arXiv:1902.04232} \BibitemShut {NoStop}%
\bibitem [{\citenamefont {Lobo}\ \emph {et~al.}(2021)\citenamefont {Lobo},
  \citenamefont {Rodrigues}, \citenamefont {de~S.~Silva}, \citenamefont
  {Simpson},\ and\ \citenamefont {Visser}}]{lobo_novel_2021}%
  \BibitemOpen
  \bibfield  {author} {\bibinfo {author} {\bibfnamefont {F.~S.~N.}\
  \bibnamefont {Lobo}}, \bibinfo {author} {\bibfnamefont {M.~E.}\ \bibnamefont
  {Rodrigues}}, \bibinfo {author} {\bibfnamefont {M.~V.}\ \bibnamefont
  {de~S.~Silva}}, \bibinfo {author} {\bibfnamefont {A.}~\bibnamefont
  {Simpson}},\ and\ \bibinfo {author} {\bibfnamefont {M.}~\bibnamefont
  {Visser}},\ }\bibfield  {title} {\bibinfo {title} {{Novel Black-Bounce
  Spacetimes: Wormholes, Regularity, Energy Conditions, and Causal
  Structure}},\ }\href {https://doi.org/10.1103/PhysRevD.103.084052} {\bibfield
   {journal} {\bibinfo  {journal} {Phys. Rev. D}\ }\textbf {\bibinfo {volume}
  {103}},\ \bibinfo {pages} {084052} (\bibinfo {year} {2021})},\ \Eprint
  {https://arxiv.org/abs/2009.12057} {arXiv:2009.12057} \BibitemShut {NoStop}%
\bibitem [{\citenamefont {Franzin}\ \emph
  {et~al.}(2021{\natexlab{b}})\citenamefont {Franzin}, \citenamefont
  {Liberati}, \citenamefont {Mazza}, \citenamefont {Simpson},\ and\
  \citenamefont {Visser}}]{Franzin:2021pyi}%
  \BibitemOpen
  \bibfield  {author} {\bibinfo {author} {\bibfnamefont {E.}~\bibnamefont
  {Franzin}}, \bibinfo {author} {\bibfnamefont {S.}~\bibnamefont {Liberati}},
  \bibinfo {author} {\bibfnamefont {J.}~\bibnamefont {Mazza}}, \bibinfo
  {author} {\bibfnamefont {A.}~\bibnamefont {Simpson}},\ and\ \bibinfo {author}
  {\bibfnamefont {M.}~\bibnamefont {Visser}},\ }\bibfield  {title} {\bibinfo
  {title} {{Charged black-bounce spacetimes}},\ }\href
  {https://doi.org/10.1088/1475-7516/2021/07/036} {\bibfield  {journal}
  {\bibinfo  {journal} {J. Cosmol. Astropart. Phys.}\ }\textbf {\bibinfo
  {volume} {\normalfont07}}\bibfield  {number} {\bibinfo  {number} { (2021)},\
  \bibinfo {pages} {036}},\ }\Eprint {https://arxiv.org/abs/2104.11376}
  {arXiv:2104.11376 [gr-qc]} \BibitemShut {NoStop}%
\bibitem [{\citenamefont {Johannsen}(2013)}]{Johannsen:2015pca}%
  \BibitemOpen
  \bibfield  {author} {\bibinfo {author} {\bibfnamefont {T.}~\bibnamefont
  {Johannsen}},\ }\bibfield  {title} {\bibinfo {title} {{Regular Black Hole
  Metric with Three Constants of Motion}},\ }\href
  {https://doi.org/10.1103/PhysRevD.88.044002} {\bibfield  {journal} {\bibinfo
  {journal} {Phys. Rev. D}\ }\textbf {\bibinfo {volume} {88}},\ \bibinfo
  {pages} {044002} (\bibinfo {year} {2013})},\ \Eprint
  {https://arxiv.org/abs/1501.02809} {arXiv:1501.02809 [gr-qc]} \BibitemShut
  {NoStop}%
\bibitem [{\citenamefont {Johannsen}(2016)}]{Johannsen:2015mdd}%
  \BibitemOpen
  \bibfield  {author} {\bibinfo {author} {\bibfnamefont {T.}~\bibnamefont
  {Johannsen}},\ }\bibfield  {title} {\bibinfo {title} {{Sgr A* and General
  Relativity}},\ }\href {https://doi.org/10.1088/0264-9381/33/11/113001}
  {\bibfield  {journal} {\bibinfo  {journal} {Class. Quantum Grav.}\ }\textbf
  {\bibinfo {volume} {33}},\ \bibinfo {pages} {113001} (\bibinfo {year}
  {2016})},\ \Eprint {https://arxiv.org/abs/1512.03818} {arXiv:1512.03818
  [astro-ph.GA]} \BibitemShut {NoStop}%
\bibitem [{\citenamefont {Berti}\ \emph {et~al.}(2006)\citenamefont {Berti},
  \citenamefont {Cardoso},\ and\ \citenamefont {Casals}}]{Berti:2005gp}%
  \BibitemOpen
  \bibfield  {author} {\bibinfo {author} {\bibfnamefont {E.}~\bibnamefont
  {Berti}}, \bibinfo {author} {\bibfnamefont {V.}~\bibnamefont {Cardoso}},\
  and\ \bibinfo {author} {\bibfnamefont {M.}~\bibnamefont {Casals}},\
  }\bibfield  {title} {\bibinfo {title} {{Eigenvalues and eigenfunctions of
  spin-weighted spheroidal harmonics in four and higher dimensions}},\ }\href
  {https://doi.org/10.1103/PhysRevD.73.109902} {\bibfield  {journal} {\bibinfo
  {journal} {Phys. Rev. D}\ }\textbf {\bibinfo {volume} {73}},\ \bibinfo
  {pages} {024013} (\bibinfo {year} {2006})},\ \bibinfo {note} {[Erratum: Phys.
  Rev. D {\bf73}, 109902(E) (2006)]},\ \Eprint
  {https://arxiv.org/abs/gr-qc/0511111} {arXiv:gr-qc/0511111} \BibitemShut
  {NoStop}%
\bibitem [{\citenamefont {Leaver}(1985)}]{Leaver:1985ax}%
  \BibitemOpen
  \bibfield  {author} {\bibinfo {author} {\bibfnamefont {E.~W.}\ \bibnamefont
  {Leaver}},\ }\bibfield  {title} {\bibinfo {title} {{An analytic
  representation for the quasi normal modes of Kerr black holes}},\ }\href
  {https://doi.org/10.1098/rspa.1985.0119} {\bibfield  {journal} {\bibinfo
  {journal} {Proc. Roy. Soc. Lond. A}\ }\textbf {\bibinfo {volume} {402}},\
  \bibinfo {pages} {285} (\bibinfo {year} {1985})}\BibitemShut {NoStop}%
\bibitem [{\citenamefont {Schutz}\ and\ \citenamefont
  {Will}(1985)}]{Schutz:1985km}%
  \BibitemOpen
  \bibfield  {author} {\bibinfo {author} {\bibfnamefont {B.~F.}\ \bibnamefont
  {Schutz}}\ and\ \bibinfo {author} {\bibfnamefont {C.~M.}\ \bibnamefont
  {Will}},\ }\bibfield  {title} {\bibinfo {title} {{Black hole normal modes: A
  semianalytic approach}},\ }\href {https://doi.org/10.1086/184453} {\bibfield
  {journal} {\bibinfo  {journal} {Astrophys. J. Lett.}\ }\textbf {\bibinfo
  {volume} {291}},\ \bibinfo {pages} {L33} (\bibinfo {year}
  {1985})}\BibitemShut {NoStop}%
\bibitem [{\citenamefont {Iyer}\ and\ \citenamefont
  {Will}(1987)}]{Iyer:1986np}%
  \BibitemOpen
  \bibfield  {author} {\bibinfo {author} {\bibfnamefont {S.}~\bibnamefont
  {Iyer}}\ and\ \bibinfo {author} {\bibfnamefont {C.~M.}\ \bibnamefont
  {Will}},\ }\bibfield  {title} {\bibinfo {title} {{Black Hole Normal Modes: A
  WKB Approach. I. Foundations and Application of a Higher Order WKB Analysis
  of Potential Barrier Scattering}},\ }\href
  {https://doi.org/10.1103/PhysRevD.35.3621} {\bibfield  {journal} {\bibinfo
  {journal} {Phys. Rev. D}\ }\textbf {\bibinfo {volume} {35}},\ \bibinfo
  {pages} {3621} (\bibinfo {year} {1987})}\BibitemShut {NoStop}%
\bibitem [{\citenamefont {Seidel}\ and\ \citenamefont
  {Iyer}(1990)}]{Seidel:1989bp}%
  \BibitemOpen
  \bibfield  {author} {\bibinfo {author} {\bibfnamefont {E.}~\bibnamefont
  {Seidel}}\ and\ \bibinfo {author} {\bibfnamefont {S.}~\bibnamefont {Iyer}},\
  }\bibfield  {title} {\bibinfo {title} {{Black hole normal modes: A WKB
  approach. IV. Kerr black holes}},\ }\href
  {https://doi.org/10.1103/PhysRevD.41.374} {\bibfield  {journal} {\bibinfo
  {journal} {Phys. Rev. D}\ }\textbf {\bibinfo {volume} {41}},\ \bibinfo
  {pages} {374} (\bibinfo {year} {1990})}\BibitemShut {NoStop}%
\bibitem [{\citenamefont {Kokkotas}(1991)}]{Kokkotas:1991vz}%
  \BibitemOpen
  \bibfield  {author} {\bibinfo {author} {\bibfnamefont {K.~D.}\ \bibnamefont
  {Kokkotas}},\ }\bibfield  {title} {\bibinfo {title} {{Normal modes of the
  Kerr black hole}},\ }\href {https://doi.org/10.1088/0264-9381/8/12/006}
  {\bibfield  {journal} {\bibinfo  {journal} {Class. Quantum Grav.}\ }\textbf
  {\bibinfo {volume} {8}},\ \bibinfo {pages} {2217} (\bibinfo {year}
  {1991})}\BibitemShut {NoStop}%
\bibitem [{\citenamefont {Chandrasekhar}\ and\ \citenamefont
  {Detweiler}(1975)}]{Chandrasekhar:1975zza}%
  \BibitemOpen
  \bibfield  {author} {\bibinfo {author} {\bibfnamefont {S.}~\bibnamefont
  {Chandrasekhar}}\ and\ \bibinfo {author} {\bibfnamefont {S.~L.}\ \bibnamefont
  {Detweiler}},\ }\bibfield  {title} {\bibinfo {title} {{The quasi-normal modes
  of the Schwarzschild black hole}},\ }\href
  {https://doi.org/10.1098/rspa.1975.0112} {\bibfield  {journal} {\bibinfo
  {journal} {Proc. Roy. Soc. Lond. A}\ }\textbf {\bibinfo {volume} {344}},\
  \bibinfo {pages} {441} (\bibinfo {year} {1975})}\BibitemShut {NoStop}%
\bibitem [{\citenamefont {Iyer}(1987)}]{Iyer:1986nq}%
  \BibitemOpen
  \bibfield  {author} {\bibinfo {author} {\bibfnamefont {S.}~\bibnamefont
  {Iyer}},\ }\bibfield  {title} {\bibinfo {title} {{Black Hole Normal Modes: A
  WKB Approach. II. Schwarzschild black holes}},\ }\href
  {https://doi.org/10.1103/PhysRevD.35.3632} {\bibfield  {journal} {\bibinfo
  {journal} {Phys. Rev. D}\ }\textbf {\bibinfo {volume} {35}},\ \bibinfo
  {pages} {3632} (\bibinfo {year} {1987})}\BibitemShut {NoStop}%
\bibitem [{\citenamefont {Konoplya}(2003)}]{Konoplya:2003ii}%
  \BibitemOpen
  \bibfield  {author} {\bibinfo {author} {\bibfnamefont {R.~A.}\ \bibnamefont
  {Konoplya}},\ }\bibfield  {title} {\bibinfo {title} {{Quasinormal behavior of
  the D-dimensional Schwarzschild black hole and higher order WKB approach}},\
  }\href {https://doi.org/10.1103/PhysRevD.68.024018} {\bibfield  {journal}
  {\bibinfo  {journal} {Phys. Rev. D}\ }\textbf {\bibinfo {volume} {68}},\
  \bibinfo {pages} {024018} (\bibinfo {year} {2003})},\ \Eprint
  {https://arxiv.org/abs/gr-qc/0303052} {arXiv:gr-qc/0303052} \BibitemShut
  {NoStop}%
\bibitem [{\citenamefont {Matyjasek}\ and\ \citenamefont
  {Opala}(2017)}]{Matyjasek:2017psv}%
  \BibitemOpen
  \bibfield  {author} {\bibinfo {author} {\bibfnamefont {J.}~\bibnamefont
  {Matyjasek}}\ and\ \bibinfo {author} {\bibfnamefont {M.}~\bibnamefont
  {Opala}},\ }\bibfield  {title} {\bibinfo {title} {{Quasinormal modes of black
  holes. The improved semianalytic approach}},\ }\href
  {https://doi.org/10.1103/PhysRevD.96.024011} {\bibfield  {journal} {\bibinfo
  {journal} {Phys. Rev. D}\ }\textbf {\bibinfo {volume} {96}},\ \bibinfo
  {pages} {024011} (\bibinfo {year} {2017})},\ \Eprint
  {https://arxiv.org/abs/1704.00361} {arXiv:1704.00361 [gr-qc]} \BibitemShut
  {NoStop}%
\bibitem [{\citenamefont {Konoplya}\ \emph {et~al.}(2019)\citenamefont
  {Konoplya}, \citenamefont {Zhidenko},\ and\ \citenamefont
  {Zinhailo}}]{Konoplya:2019hlu}%
  \BibitemOpen
  \bibfield  {author} {\bibinfo {author} {\bibfnamefont {R.~A.}\ \bibnamefont
  {Konoplya}}, \bibinfo {author} {\bibfnamefont {A.}~\bibnamefont {Zhidenko}},\
  and\ \bibinfo {author} {\bibfnamefont {A.~F.}\ \bibnamefont {Zinhailo}},\
  }\bibfield  {title} {\bibinfo {title} {{Higher order WKB formula for
  quasinormal modes and grey-body factors: recipes for quick and accurate
  calculations}},\ }\href {https://doi.org/10.1088/1361-6382/ab2e25} {\bibfield
   {journal} {\bibinfo  {journal} {Class. Quantum Grav.}\ }\textbf {\bibinfo
  {volume} {36}},\ \bibinfo {pages} {155002} (\bibinfo {year} {2019})},\
  \Eprint {https://arxiv.org/abs/1904.10333} {arXiv:1904.10333 [gr-qc]}
  \BibitemShut {NoStop}%
\bibitem [{\citenamefont {Churilova}\ and\ \citenamefont
  {Stuchlik}(2020)}]{Churilova:2019cyt}%
  \BibitemOpen
  \bibfield  {author} {\bibinfo {author} {\bibfnamefont {M.~S.}\ \bibnamefont
  {Churilova}}\ and\ \bibinfo {author} {\bibfnamefont {Z.}~\bibnamefont
  {Stuchlik}},\ }\bibfield  {title} {\bibinfo {title} {{Ringing of the regular
  black-hole/wormhole transition}},\ }\href
  {https://doi.org/10.1088/1361-6382/ab7717} {\bibfield  {journal} {\bibinfo
  {journal} {Class. Quantum Grav.}\ }\textbf {\bibinfo {volume} {37}},\
  \bibinfo {pages} {075014} (\bibinfo {year} {2020})},\ \Eprint
  {https://arxiv.org/abs/1911.11823} {arXiv:1911.11823 [gr-qc]} \BibitemShut
  {NoStop}%
\bibitem [{\citenamefont {Konoplya}\ and\ \citenamefont
  {Zhidenko}(2010)}]{konoplya_passage_2010}%
  \BibitemOpen
  \bibfield  {author} {\bibinfo {author} {\bibfnamefont {R.~A.}\ \bibnamefont
  {Konoplya}}\ and\ \bibinfo {author} {\bibfnamefont {A.}~\bibnamefont
  {Zhidenko}},\ }\bibfield  {title} {\bibinfo {title} {{Passage of Radiation
  through Wormholes of Arbitrary Shape}},\ }\href
  {https://doi.org/10.1103/PhysRevD.81.124036} {\bibfield  {journal} {\bibinfo
  {journal} {Phys. Rev. D}\ }\textbf {\bibinfo {volume} {81}},\ \bibinfo
  {pages} {124036} (\bibinfo {year} {2010})},\ \Eprint
  {https://arxiv.org/abs/1004.1284} {arXiv:1004.1284} \BibitemShut {NoStop}%
\bibitem [{\citenamefont {Penrose}\ and\ \citenamefont
  {Floyd}(1971)}]{penrose_extraction_1971}%
  \BibitemOpen
  \bibfield  {author} {\bibinfo {author} {\bibfnamefont {R.}~\bibnamefont
  {Penrose}}\ and\ \bibinfo {author} {\bibfnamefont {R.~M.}\ \bibnamefont
  {Floyd}},\ }\bibfield  {title} {\bibinfo {title} {Extraction of {{Rotational
  Energy}} from a {{Black Hole}}},\ }\href
  {https://doi.org/10.1038/physci229177a0} {\bibfield  {journal} {\bibinfo
  {journal} {Nat. Phys. Sci.}\ }\textbf {\bibinfo {volume} {229}},\ \bibinfo
  {pages} {177} (\bibinfo {year} {1971})}\BibitemShut {NoStop}%
\bibitem [{\citenamefont
  {Chandrasekhar}(1983)}]{chandrasekhar_mathematical_1983}%
  \BibitemOpen
  \bibfield  {author} {\bibinfo {author} {\bibfnamefont {S.}~\bibnamefont
  {Chandrasekhar}},\ }\href@noop {} {\emph {\bibinfo {title} {The
  {{Mathematical Theory}} of {{Black Holes}}}}}\ (\bibinfo  {publisher}
  {{Oxford University Press}},\ \bibinfo {address} {{Walton Street, Oxford}},\
  \bibinfo {year} {1983})\BibitemShut {NoStop}%
\bibitem [{\citenamefont {Bardeen}\ \emph {et~al.}(1972)\citenamefont
  {Bardeen}, \citenamefont {Press},\ and\ \citenamefont
  {Teukolsky}}]{bardeen_rotating_1972}%
  \BibitemOpen
  \bibfield  {author} {\bibinfo {author} {\bibfnamefont {J.~M.}\ \bibnamefont
  {Bardeen}}, \bibinfo {author} {\bibfnamefont {W.~H.}\ \bibnamefont {Press}},\
  and\ \bibinfo {author} {\bibfnamefont {S.~A.}\ \bibnamefont {Teukolsky}},\
  }\bibfield  {title} {\bibinfo {title} {Rotating {{Black Holes}}: {{Locally
  Nonrotating Frames}}, {{Energy Extraction}}, and {{Scalar Synchrotron
  Radiation}}},\ }\href {https://doi.org/10.1086/151796} {\bibfield  {journal}
  {\bibinfo  {journal} {The Astrophysical Journal}\ }\textbf {\bibinfo {volume}
  {178}},\ \bibinfo {pages} {347} (\bibinfo {year} {1972})}\BibitemShut
  {NoStop}%
\bibitem [{\citenamefont {Wald}(1974)}]{wald_energy_1974}%
  \BibitemOpen
  \bibfield  {author} {\bibinfo {author} {\bibfnamefont {R.~M.}\ \bibnamefont
  {Wald}},\ }\bibfield  {title} {\bibinfo {title} {Energy {{Limits}} on the
  {{Penrose Process}}},\ }\href {https://doi.org/10.1086/152959} {\bibfield
  {journal} {\bibinfo  {journal} {Astrophys. J.}\ }\textbf {\bibinfo {volume}
  {191}},\ \bibinfo {pages} {231} (\bibinfo {year} {1974})}\BibitemShut
  {NoStop}%
\bibitem [{\citenamefont {Kovetz}\ and\ \citenamefont
  {Piran}(1975)}]{kovetz_efficiency_1975}%
  \BibitemOpen
  \bibfield  {author} {\bibinfo {author} {\bibfnamefont {A.}~\bibnamefont
  {Kovetz}}\ and\ \bibinfo {author} {\bibfnamefont {T.}~\bibnamefont {Piran}},\
  }\bibfield  {title} {\bibinfo {title} {The efficiency of the {{Penrose}}
  process},\ }\href {https://doi.org/10.1007/BF02813727,10.1007/BF02813753}
  {\bibfield  {journal} {\bibinfo  {journal} {Lett. Nuovo Cimento}\ }\textbf
  {\bibinfo {volume} {12}},\ \bibinfo {pages} {39} (\bibinfo {year} {1975})},\
  \bibinfo {note} {[Erratum: Lett. Nuovo Cimento {\bf14}, 560
  (1975)]}\BibitemShut {NoStop}%
\bibitem [{\citenamefont {Teukolsky}\ and\ \citenamefont
  {Press}(1974)}]{Teukolsky:1974yv}%
  \BibitemOpen
  \bibfield  {author} {\bibinfo {author} {\bibfnamefont {S.~A.}\ \bibnamefont
  {Teukolsky}}\ and\ \bibinfo {author} {\bibfnamefont {W.~H.}\ \bibnamefont
  {Press}},\ }\bibfield  {title} {\bibinfo {title} {{Perturbations of a
  rotating black hole. III\@. Interaction of the hole with gravitational and
  electromagnetic radiation}},\ }\href {https://doi.org/10.1086/153180}
  {\bibfield  {journal} {\bibinfo  {journal} {Astrophys. J.}\ }\textbf
  {\bibinfo {volume} {193}},\ \bibinfo {pages} {443} (\bibinfo {year}
  {1974})}\BibitemShut {NoStop}%
\bibitem [{\citenamefont {Lima~Junior}\ \emph {et~al.}(2020)\citenamefont
  {Lima~Junior}, \citenamefont {Benone},\ and\ \citenamefont
  {Crispino}}]{Junior:2020lse}%
  \BibitemOpen
  \bibfield  {author} {\bibinfo {author} {\bibfnamefont {H.~C.~D.}\
  \bibnamefont {Lima~Junior}}, \bibinfo {author} {\bibfnamefont {C.~L.}\
  \bibnamefont {Benone}},\ and\ \bibinfo {author} {\bibfnamefont {L.~C.~B.}\
  \bibnamefont {Crispino}},\ }\bibfield  {title} {\bibinfo {title} {{Scalar
  absorption: Black holes versus wormholes}},\ }\href
  {https://doi.org/10.1103/PhysRevD.101.124009} {\bibfield  {journal} {\bibinfo
   {journal} {Phys. Rev. D}\ }\textbf {\bibinfo {volume} {101}},\ \bibinfo
  {pages} {124009} (\bibinfo {year} {2020})},\ \Eprint
  {https://arxiv.org/abs/2006.03967} {arXiv:2006.03967 [gr-qc]} \BibitemShut
  {NoStop}%
\bibitem [{\citenamefont {Pelavas}\ \emph {et~al.}(2001)\citenamefont
  {Pelavas}, \citenamefont {Neary},\ and\ \citenamefont
  {Lake}}]{pelavas_properties_2001}%
  \BibitemOpen
  \bibfield  {author} {\bibinfo {author} {\bibfnamefont {N.}~\bibnamefont
  {Pelavas}}, \bibinfo {author} {\bibfnamefont {N.}~\bibnamefont {Neary}},\
  and\ \bibinfo {author} {\bibfnamefont {K.}~\bibnamefont {Lake}},\ }\bibfield
  {title} {\bibinfo {title} {{Properties of the Instantaneous Ergo Surface of a
  Kerr Black Hole}},\ }\href {https://doi.org/10.1088/0264-9381/18/7/314}
  {\bibfield  {journal} {\bibinfo  {journal} {Class. Quantum Grav.}\ }\textbf
  {\bibinfo {volume} {18}},\ \bibinfo {pages} {1319} (\bibinfo {year}
  {2001})},\ \Eprint {https://arxiv.org/abs/gr-qc/0012052}
  {arXiv:gr-qc/0012052} \BibitemShut {NoStop}%
\bibitem [{\citenamefont {Bender}\ and\ \citenamefont
  {Orszag}(1999)}]{bender_advanced_1999}%
  \BibitemOpen
  \bibfield  {author} {\bibinfo {author} {\bibfnamefont {C.~M.}\ \bibnamefont
  {Bender}}\ and\ \bibinfo {author} {\bibfnamefont {S.~A.}\ \bibnamefont
  {Orszag}},\ }\href@noop {} {\emph {\bibinfo {title} {Advanced {{Mathematical
  Methods}} for {{Scientists}} and {{Engineers I}}}}},\ \bibinfo {edition}
  {1st}\ ed.\ (\bibinfo  {publisher} {{Springer, New York, NY}},\ \bibinfo
  {year} {1999})\BibitemShut {NoStop}%
\end{thebibliography}%

\end{document}